**BUPD: A Bayesian under-parameterized basket design with the unit information prior in oncology trials**

Ryo Kitabayashi[a], Hiroyuki Sato[a], and Akihiro Hirakawa[a,*]

[a]*Department of Clinical Biostatistics, Graduate School of Medical and Dental Sciences, Institute of Science Tokyo, 1-5-45 Yushima, Bunkyo-ku, Tokyo 113-8510, Japan*

**\*Corresponding author:**
Akihiro Hirakawa, PhD
Department of Clinical Biostatistics, Graduate School of Medical and Dental Sciences, Institute of Science Tokyo, 1-5-45 Yushima, Bunkyo-ku, Tokyo 113-8510, Japan
E-mail: a-hirakawa.crc@tmd.ac.jp

**Abstract**

Basket trials in oncology enroll multiple patients with cancer harboring identical gene alterations and evaluate their response to targeted therapies across cancer types. Several existing methods have extended a Bayesian hierarchical model borrowing information on the response rates in different cancer types to account for the heterogeneity of drug effects. However, these methods rely on several pre-specified parameters to account for the heterogeneity of response rates among different cancer types. Here, we propose a novel Bayesian under-parameterized basket design with a unit information prior (BUPD) that uses only one (or two) pre-specified parameters to control the amount of information borrowed among cancer types, considering the heterogeneity of response rates. BUPD adapts the unit information prior approach, originally developed for borrowing information from historical clinical trial data, to enable mutual information borrowing between two cancer types. BUPD enables flexible controls of the type 1 error rate and power by explicitly specifying the strength of borrowing while providing interpretable estimations of response rates. Simulation studies revealed that BUPD reduced the type 1 error rate in scenarios with few ineffective cancer types and improved the power in scenarios with few effective cancer types better than five existing methods. This study also illustrated the efficiency of BUPD using response rates from a real basket trial.

# 1. Introduction

Targeted therapy has gained attention as an alternative to conventional chemotherapy in oncology owing to advances in genome sequencing technologies. Basket trials, which enroll multiple patients with cancer with identical gene alterations and evaluate the efficacy (e.g., response rate) of a targeted therapy across cancer types, have been used for early-phase clinical trials.[1–3] Basket trials often reveal that the efficacy of the targeted therapy varies depending on the cancer type, as demonstrated by the trial of vemurafenib for non-melanoma cancers with *BRAF*-V600 mutations.[4] Due of the limited sample size for each cancer type, development of efficient statistical methods is necessary to evaluate the response rates while accounting for their heterogeneity.

The Bayesian hierarchical model (BHM),[5,6] widely used in early-phase basket trials, enables the borrowing of information on response rates among cancer types by assuming exchangeability and improves the power for declaring the drug efficacy of each cancer type over methods evaluating the efficacy of each cancer type independently. However, BHM can significantly inflate the type 1 error rate when the true response rate for each cancer type is heterogeneous.[7] To address the heterogeneity of response rates among cancer types, numerous extensions of BHM have been proposed. These includes methods assuming the exchangeability and non-exchangeability (EXNEX) model for different cancer types,[8–10] using the calibrated variance parameter for the prior treatment effect,[11] incorporating the similarity of response rates among cancer types into BHM,[12] classifying the cancer types into two or more clusters,[13–15] and using the Bayesian hypothesis testing framework.[16,17] However, these methods rely on several pre-specified parameters (e.g., ten in EXNEX among them) to account for the heterogeneity of response rates among cancer types. This presents a challenge in identifying optimal parameter specifications that minimize the inflation of the type 1 error rate for each cancer type while improving statistical power, given the numerous possible combinations of pre-specified parameter values. Despite the increased complexity of these models, their improvement in the power of declaring drug efficacy compared with the original BHM is limited.[18] Several methods using a model with few pre-specified parameters have been proposed.[19–21] Our simulation studies have demonstrated that these methods can lead to extremely strong borrowing of information even when a using non-informative prior or parameter specification recommended in the original article, resulting in reduced power in scenarios with few effective cancer types (see Section 3).

In this study, we have developed a new Bayesian under-parameterized basket design with a unit information prior (BUPD) that uses only one (or two) pre-specified parameters to control the amount of information borrowed among cancer types. We formulated BUPD by extending the conceptual framework of the unit information prior approach introduced by Jin and Yin (2021).[22] While their approach is designed for unidirectional information borrowing from multiple historical clinical trial datasets into the current trial, we introduce a novel methodology that enables bidirectional information borrowing between cancer types within the basket trials. Specifically, BUPD incorporates mutual borrowing based on the consistency of response rates between two cancer types, ensuring that the similarity from cancer type $i$ to $j$ is symmetric with the similarity from cancer type $j$ to $i$. This symmetry, which we designed specifically, distinguishes BUPD from previous approaches and was not addressed in the original unit information prior framework. This mechanism allows BUPD to leverage data across cancer types more effectively, ensuring the interpretability and robustness of the inference.

In BUPD, we define the unit information as the amount of information for a single patient and determine how much patient information is borrowed across cancer types when specifying the prior distribution of the response rate of each cancer type. BUPD introduces two novel parameters to enhance the flexibility and interpretability of information borrowing among different cancer types. The first parameter quantifies the overall strength of information borrowing, effectively representing the sample size of the total amount of information borrowed across all possible pairwise combinations of cancer types. The second parameter is a vector that accounts for the heterogeneity of the response rates by explicitly representing the consistency of each pairwise combination of cancer types. These parameters offer several advantages. Specifically, the first parameter allows flexible controls of the type 1 error rate and statistical power through a single pre-specified value, thus enabling precise adjustments to the strength of information borrowing. Moreover, because this parameter is interpretable as the sample size, BUPD provides a transparent and intuitive mechanism for controlling information borrowing. This feature prevents the estimation process from being perceived as a "black box," thereby fostering greater understanding among both statisticians and non-statisticians.

We also compared the operating characteristics of the proposed methods with those of five existing methods and assessed the parameter specifications of the proposed methods through comprehensive simulation studies. Application of the proposed methods using a real basket trial for vemurafenib has been illustrated.

In the remainder of this paper, we formulate the quantitative information borrowing among cancer types in BUPD in Section 2, compare the operating characteristics of the proposed method with those of five existing methods via a simulation study and perform a sensitivity analysis for parameter specifications in Section 3, demonstrate an application of the proposed method using real data from a basket trial in Section 4, and discuss the proposed method in detail in Section 5.

## 2. Methods

We consider a single-arm phase II basket trial. We denote the number of enrolled patients, the number of responders, and the true response rate in cancer type $i$ ($i = 1, \cdots, I$) as $n_i$, $x_i$, and $\pi_i$ respectively, and assume $x_i \sim Bin(n_i, \pi_i)$. The total sample size planned for the trial is denoted by $N = \sum_{i=1}^{I} n_i$.

## 2.1. Unit information prior for response rate

We estimate the posterior distribution of true response rate of $\pi_i$ for cancer type $i$ by information borrowing quantified based on a beta prior distribution with parameters $\alpha_i$ and $\beta_i$, i.e., $\pi_i \sim Beta(\alpha_i, \beta_i)$, which is constructed from the data of response rates on remaining cancer types $j \ (\neq i)$. Therefore, the mean $\mu_i$ and variance $\eta_i^2$ of the prior distribution for $\pi_i$ are represented as $\alpha_i/(\alpha_i + \beta_i)$ and $\alpha_i\beta_i/(\alpha_i + \beta_i)^2(\alpha_i + \beta_i + 1)$, respectively. To determine $\alpha_i$ and $\beta_i$, it is vital to parameterize (i) how similar the two cancer types $i$ and $j$ are and (ii) how much total amount of information is borrowed throughout all pairwise combination.

To achieve the parameterization (i), the consistency of response rate between cancer types $i$ and $j$ is measured by introducing a weight parameter $w_{ij} \in (0, 0.5)$ conditional on $w_{ij} = w_{ji}$ (i.e., a symmetric weight between two cancer types) and $\sum_{i=1}^{I} \sum_{j \neq i} w_{ij} = 1$. We defined the mean of the prior distribution for $\pi_i$ in cancer type $i$ as the weighted mean of the maximum likelihood estimators of the response rate in other cancer types.

$$\mu_i = \sum_{j \neq i} \frac{w_{ij}}{\sum_{j \neq i} w_{ij}} \pi_j. \tag{1}$$

Let $\boldsymbol{w}$ denotes the weight parameter vector comprising the $I(I-1)$ weight parameters where $i \neq j$, i.e., $\boldsymbol{w} = (w_{12}, \cdots, w_{1I}, w_{21}, \cdots, w_{I(I-1)})^T$. The parameter $\boldsymbol{w}$ is calculated based on the observed data or modeled using a hyper-prior, as described in Section 2.2.

Next, we extend the definition of unit information proposed by Jin and Yin (2021)[22] to basket trials. We define the unit information for cancer type $i$ as the Fisher information of $\pi_i$ per one patient, that is,

$$UI(\pi_i) = \frac{1}{n_i} E\left[-\frac{\partial^2 \log L(\pi_i | x_i, n_i)}{\partial \pi_i^2}\right] = \frac{1}{\pi_i(1 - \pi_i)}, \tag{2}$$

where $L(\pi_i | x_i, n_i)$ is the likelihood function of the binomial distribution, i.e., $L(\pi_i | x_i, n_i) \propto \pi_i^{x_i}(1 - \pi_i)^{n_i - x_i}$, and $-\partial^2 \log L(\pi_i | x_i, n_i)/\partial \pi_i^2$ is the Fisher information of $\pi_i$. Using the weight parameter $w_{ij}$, the information that cancer type $i$ borrows from other cancer types for a single patient is quantified by $\sum_{j \neq i} w_{ij} UI(\pi_j)$. To achieve parameterization (ii), we introduced the parameter of $M \ (> 0)$ as the sample size of the total amount of information borrowed throughout all pairwise combinations of $I(I-1)/2$ between cancer types $i$ and $j$. The total amount of information borrowed from other cancer types for cancer type $i$ is represented as

$$M \sum_{j \neq i} w_{ij} UI(\pi_j). \tag{3}$$

According to the original article of unit information prior[22], we define the variance of the prior distribution for $\pi_i$ using Equation (3) as follows,

$$\eta_i^2 = \left\{ M \sum_{j \neq i} w_{ij} UI(\pi_j) \right\}^{-1}. \tag{4}$$

The parameter $M$ is a pre-specified value or modeled using a hyper-prior, as described in Section 2.2. The mean, $\mu_i$, and the variance, $\eta_i^2$, for the prior distribution of $\pi_i$ allow the effective sample size of the posterior distribution for $\pi_i$ to be represented by a simple formula using $M$ and $w_{ij}$ (see Section 2.3).

A higher (or lower) value of $w_{ij}$ indicates that the response rates of cancer types $i$ and $j$ have relatively higher (or lower) consistency than those of the other two cancer types with lower values of the weight parameter. Therefore, we can control the heterogeneity of response rates among cancer types using the weight parameter $w_{ij}$. As the value of $M$ approaches zero, the variance $\eta_i^2$ of the prior distribution for $\pi_i$ in Equation (4) becomes larger (i.e., amount of information contained in the prior distribution of $\pi_i$ approaches zero), resulting in weaker information borrowing among cancer types. We can also explicitly control the total strength of information borrowing among cancer types using parameter $M$. In the special case where cancer type $i$ borrows equal information of $n$ patients from the other cancer types under $n_1 = \cdots = n_I = n$, the value of $M$ is equal to $nI(I-1)$. Furthermore, $Mw_{ij}$ represents the sample size of information mutually borrowed between cancer types $i$ and $j$, providing quantitative information on the number of patients of each cancer type from which information was borrowed.

Because the true response rate $\pi_j$ for cancer type $j$, used in Equations (1) and (4) for $\mu_i$ and $\eta_i^2$, is unknown, we have used the maximum likelihood estimator $\hat{\pi}_j = x_j/n_j$, similar to Jin and Yin (2021).[22] The estimates is based on the observed data after trial initiation and is used to model the prior distribution of $\pi_i$. Using $\mu_i$ and $\eta_i^2$ derived from $\hat{\pi}_j$, along with the two parameters $w$ and $M$, we solve $\mu_i = \alpha_i/(\alpha_i + \beta_i)$ and $\eta_i^2 = \alpha_i\beta_i/(\alpha_i + \beta_i)^2(\alpha_i + \beta_i + 1)$ with respect to $\alpha_i$ and $\beta_i$, obtaining the two parameters of $Beta(\alpha_i, \beta_i)$ as follows:

$$\alpha_i = \mu_i \left\{ \frac{\mu_i(1-\mu_i)}{\eta_i^2} - 1 \right\}, \beta_i = (1-\mu_i) \left\{ \frac{\mu_i(1-\mu_i)}{\eta_i^2} - 1 \right\}. \tag{5}$$

An extremely small $\hat{\pi}_j$ leads to the inflation of $UI(\hat{\pi}_j)$, $\alpha_i$, and $\beta_i$, resulting in the posterior response rate estimation for cancer type $i$ being dominated by information from other cancer types. To address this, we replace $UI(\hat{\pi}_j)$ with $UI(0.05)$ when $\hat{\pi}_j < 0.05$. This adjustment is justified because 0.05 represents the minimum response rate commonly used in oncology clinical trials (e.g., the null response rate for cancer types lacking a standard of care), and an observed extremely small response rate can be interpreted as an estimate of the parameter with this minimum response rate as its true value.

### *2.2. Settings of two parameters for information borrowing*

In this section, we propose several specifications on the two parameters of $\boldsymbol{w}$ and $M$. We propose the use of Dirichlet prior as the hyper-prior for $\boldsymbol{w}$ according to $\sum_{i=1}^{I}\sum_{j \neq i} w_{ij} = 1$, termed as BUPD-D,

$$z_1, \cdots, z_{\frac{I(I-1)}{2}} \sim Dirichlet\left(\tilde{z}_1, \cdots, \tilde{z}_{\frac{I(I-1)}{2}}\right), \tag{6}$$

where $z_l = w_{ij} + w_{ji}$, $l = 1, \cdots, I(I-1)/2$ (i.e., $w_{ij} = z_l/2$). As the consistency of the response rates between the two cancer types was unknown before the trial, we recommend using the value of $\tilde{z}_l = 1$ as a non-informative Dirichlet prior.

We also propose the use of the distance between the posterior distributions of $\pi_i$ and $\pi_j$ to calculate $\boldsymbol{w}$ without its hyper-prior (BUPD-JS), and with its hyper-prior (BUPD-JSH). We calculated the Jensen–Shannon divergence $d_{ij}$ between cancer types $i$ and $j$ as follows:

$$d_{ij} = \frac{1}{2}\left\{\int f(\pi|n_i, x_i) \log \frac{f(\pi|n_i, x_i)}{f(\pi|n_j, x_j)} d\pi + \int f(\pi|n_j, x_j) \log \frac{f(\pi|n_j, x_j)}{f(\pi|n_i, x_i)} d\pi\right\}, \tag{7}$$

where $f(\pi|n_i, x_i)$ is the probability density function of $Beta(1 + x_i, 1 + (n_i - x_i))$ obtained by assuming the prior distribution of $Beta(1,1)$. Using distance $d_{ij}$, we define $w_{ij}$s as

$$w_{ij} = \frac{\exp(-d_{ij})}{\sum_{i=1}^{I}\sum_{j \neq i} \exp(-d_{ij})}. \tag{8}$$

We termed the model using Equation (8) to calculate $\boldsymbol{w}$ as BUPD-JS.

Several methods for quantifying the weight parameter of $w_{ij}$ based on the distance between the two posterior distributions of the response rates have been proposed.[12,21,23] However, the specific methods, including BUPD-JS, improving the posterior estimation of the response rates remain unknown. Furthermore, accuracy of the distance estimation is limited to a small number of patients. The formula for transposing $d_{ij}$ to $w_{ij}$ should possess the following properties: all $w_{ij}$s should have similar values regardless of the values of $d_{ij}$ when homogeneous response rates among cancer types are expected, and the values of $w_{ij}$ should vary depending on the values of $d_{ij}$ when heterogeneous response rates among cancer types are expected. In addition, to control the influence of $d_{ij}$ on weight $w_{ij}$, we introduce the common parameter $s$ $(> 0)$ for all $w_{ij}$s in Equation (8):

$$w_{ij} = \frac{\exp\left(-\frac{d_{ij}}{s}\right)}{\sum_{i=1}^{I} \sum_{j \neq i} \exp\left(-\frac{d_{ij}}{s}\right)}. \tag{9}$$

Let $d_{i^*j^*}$ and $w_{i^*j^*}$ denote the minimum $d_{ij}$ in all possible combinations between the two cancer types (i.e., $I(I-1)/2$ combinations) and the weight transformed from $d_{i^*j^*}$, respectively, and $d_{(-i^*j^*)}$ and $w_{(-i^*j^*)}$ denote the arbitrary $d_{ij}$ and $w_{ij}$ of the combinations of two cancer types other than cancer types $i^*$ and $j^*$, respectively. The parameter $s$ has the following two properties with respect to $w_{ij}$:

Property 1: Under $s \to \infty$, all values of $w_{ij}$ are equal, i.e., $w_{12} = \cdots = w_{(I-1)I} = 1/\{I(I-1)\}$
Property 2: Under $s \to 0$, $w_{i^*j^*} = w_{j^*i^*} = 0.5$ and $w_{(-i^*j^*)} = 0$

These properties are described in Section S1 of Supplementary Material. When $s$ is small, the value of $w_{ij}$ varies significantly, depending on the relative magnitude of each $d_{ij}$; whereas when $s$ is large, the value of $w_{ij}$ no longer depends on the value of $d_{ij}$. Notably, Equation (9) is identical to Equation (8) when $s = 1$. As the prior distribution for $s$, we recommend using the hyper-prior $s \sim G(0.01, 0.01)$ that is a non-informative gamma prior with the mean of 1, where $G(a, b)$ denotes the gamma distribution with shape parameter $a$ and inverse scale parameter $b$. We termed the hierarchical model using Equation (9) to calculate $\boldsymbol{w}$ with the hyper-prior for $s$ as BUPD-JSH.

To accommodate the uncertainty for the total amount of the information borrowing, we assume a uniform prior $M \sim Uniform(0, \widetilde{M})$. Consequently, the joint probability density function of the posterior distribution of $\boldsymbol{\pi} = (\pi_1, \cdots, \pi_I)^T$ in BUPD-D is obtained as follows:

$$f(\boldsymbol{\pi}|\boldsymbol{n}, \boldsymbol{x}) = \int \int \left\{ \prod_{j=1}^{I} L(\pi_j|n_j, x_j) g_{\pi_j}(\pi_j|\boldsymbol{z}, M, \widehat{\boldsymbol{\pi}}_{(-j)}) \right\} g_{\boldsymbol{z}}(\boldsymbol{z}) g_M(M) dM \, d\boldsymbol{z}, \quad (10)$$

where $\boldsymbol{n} = (n_1, n_2, \cdots, n_I)^T$, $\boldsymbol{x} = (x_1, x_2, \cdots, x_I)^T$, $\boldsymbol{z} = (z_1, \cdots, z_{I(I-1)/2})^T$, and $\widehat{\boldsymbol{\pi}}_{(-j)} = (\hat{\pi}_1, \cdots, \hat{\pi}_{j-1}, \hat{\pi}_{j+1}, \cdots, \hat{\pi}_I)^T$; and $g_{\pi_j}(\cdot)$, $g_{\boldsymbol{z}}(\cdot)$, and $g_M(\cdot)$ are the probability density functions of the prior for $\pi_j$, the hyper-prior for $\boldsymbol{z}$ and $M$, respectively. Furthermore, the probability density function of the joint posterior distribution of $\boldsymbol{\pi}$ in BUPD-JSH using Equation (9) is obtained as follows:

$$f(\boldsymbol{\pi}|\boldsymbol{n}, \boldsymbol{x}) = \int \int \left\{ \prod_{j=1}^{I} L(\pi_j|n_j, x_j) g_{\pi_j}(\pi_j|s, \boldsymbol{d}, M, \widehat{\boldsymbol{\pi}}_{(-j)}) \right\} g_s(s|\boldsymbol{d}) g_M(M) dM \, ds, (11)$$

where $\boldsymbol{d} = (d_{12}, \cdots, d_{1I}, d_{21}, \cdots, d_{I(I-1)})^T$, and $g_s(\cdot)$ is the probability density function of the hype-prior $s$. Notably, we may also consider using a fixed value of $M$ to reduce computational burden. When using a fixed for both values of $\boldsymbol{w}$ and $M$ for BUPD-JS, the posterior distribution of $\pi_i$ can be derived based on a beta distribution of $Beta\left(\hat{\alpha}_i + x_i, \hat{\beta}_i + (n_i - x_i)\right)$ based on Equation (5), leading to calculate analytically the posterior distribution without the Markov chain Monte Carlo (MCMC) method (see Section 3.1). Notably, we should use $M$ (or $\widetilde{M}) \leq N$ to aims that the amount of information for the prior distribution does not dominate that for the trial data when estimating the response rate in each cancer type.

Diagrams representing the model structures of BUPD-D, BUPD-JS, and BUPD-JSH are shown Figure S1 in the Supplemental Material. BUPD-D has the simplest model structure, as it does not include formulae for calculating $\boldsymbol{w}$, unlike BUPD-JS and BUPD-JSH. Our proposed methods are designed to require only one or two pre-specified parameters, that is, $\widetilde{M}$ and $\tilde{z}$ for BUPD-D, $M$ for BUPD-JS, and $\widetilde{M}$ and $\tilde{s}$ for BUPD-JSH.

## 2.3. *Bayesian inference*

The purpose of the trial was to declare that the drug was effective for cancer type $i$ or not, that is

$$H_0: \pi_i \leq \pi_{H_0} \text{ vs. } H_1: \pi_i \geq \pi_{H_1}, i = 1, \cdots, I,$$

where $\pi_{H_0}$ is the null response rate and $\pi_{H_1}$ is the alternative response rate for all cancer types. After estimating the posterior distribution of $\pi_i$, we declare the drug's efficacy for cancer type $i$ if the posterior response rate $\pi_i$ satisfies,

$$\Pr(\pi_i > \pi_{H_0} | \boldsymbol{n}, \boldsymbol{x}) > c, \qquad (12)$$

where $c$ represents the pre-specified cut-off value, which is often calibrated to maintain the cancer-specific type 1 error rate at the target level under the assumption of null response rates for all cancer types.

## 2.4. Effective sample size

Bayesian analysis can evaluate the amount of information included in the posterior distribution using the effective sample size (ESS). The cancer-specific posterior ESS has been discussed in some studies on basket trials.[13,19] In the proposed methods, the ESS for cancer type $i$ can be represented using fixed $\boldsymbol{w}$ and $M$ under specific conditions. When assuming that $\pi_1 = \cdots = \pi_I$ (i.e., $\hat{\pi}_1 \approx \cdots \approx \hat{\pi}_I \approx \hat{\pi}$), the posterior ESS for cancer type $i$ is approximately derived as follows:

$$\begin{aligned}
ESS_i &= n_i + \alpha_i + \beta_i \\
&= n_i + \frac{\mu_i(1-\mu_i)}{\eta_i^2} - 1 \\
&= n_i + M \left\{ \sum_{j \neq i} \frac{w_{ij}}{\sum_{j \neq i} w_{ij}} \hat{\pi}_j \right\} \left\{ \sum_{k \neq i} \frac{w_{ik}}{\sum_{k \neq i} w_{ik}} (1-\hat{\pi}_k) \right\} \left\{ \sum_{l \neq i} \frac{w_{il}}{\hat{\pi}_l(1-\hat{\pi}_l)} \right\} - 1 \\
&\approx n_i + M \frac{\hat{\pi}(1-\hat{\pi})}{\hat{\pi}(1-\hat{\pi})} \left\{ \sum_{j \neq i} \frac{w_{ij}}{\sum_{j \neq i} w_{ij}} \right\} \left\{ \sum_{k \neq i} \frac{w_{ik}}{\sum_{k \neq i} w_{ik}} \right\} \left\{ \sum_{l \neq i} w_{il} \right\} - 1 \\
&= n_i + M \sum_{l \neq i} w_{il} - 1. \qquad (13)
\end{aligned}$$

Cancer-specific posterior ESS represents the sum of the number of patients and total sample size of information borrowed from other cancer types for the cancer type of interest.

## 3. Simulation study

### 3.1. *Simulation settings*

We conducted comparative simulation studies to examine the operating characteristics of the five existing and three proposed methods. We assumed the basket trial, in which the null response rate, the alternative response rate, and the number of cancer types were $\pi_{H_0} = 0.10, \pi_{H_1} = 0.40$, and $I = 6$, respectively. The total number of patients enrolled in the basket trial was $N = \sum_{i=1}^{I} n_i = 72$, and the number of patients with each cancer type $n_i$ was determined using a multinomial distribution with equal probability (i.e., on average, $n_i = 12$).

Our proposed methods were compared with the five existing methods with different characteristics, a Beta-binomial model with no information borrowing (BBM-NB), BHM,[5,6] EXNEX,[8] multisource exchangeability model (MEM),[19] and a Beta-binomial model with the Jensen–Shannon divergence (BBM-JS).[21] The methodological descriptions and parameter specifications of the existing and proposed methods are shown in Table 1. The values of priors and hyper-priors for each method were set for non-informative priors (e.g., $\alpha_i, \beta_i$ in BBM-NB, MEM, and BBM-JS, $p_m$ in MEM, $(\delta_i^{EX1}, \delta_i^{EX}, \delta_i^{NEX})$ in EXNEX, $\tilde{z}_l$ in BUPD-D, and $\tilde{s}$ in BUPD-JSH), or set for the same values as those proposed in the original article (e.g., $\epsilon$ and $\tau$ in BBM-JSH[1]). For the parameter $\tau$ in BHM, which determines the strength of information borrowing, various hyper-prior distributions and their parameter specifications that are different from those in the original article have been discussed.[11,24–26] For ease of comparison between the existing and proposed methods, we set the hyper-prior of BHM for a gamma prior of $G(2,2)$ to achieve a power similar to that of MEM and BBM-JS under the scenario with all effective cancer types (i.e., the most preferable scenario in the trial). For $s_\tau$ in EXNEX, we selected the value that allowed EXNEX to achieve the closest power to that of BHM, MEM, and BBM-JS under the scenarios with all effective cancer types, among the candidates provided in the original article.[8] The values of $\widetilde{M}$ of $Uniform(0, \widetilde{M})$ in the three proposed methods were commonly set to the total number $N$ of patients enrolled in the basket trial.

**Table 1.** Summary and parameter specifications of the existing and proposed methods.

| Method | Summary | Parameter specifications |
|---|---|---|
| BBM-NB | Beta-binomial model with no information borrowing (BBM-NB) independently estimates the posterior $\pi_i$ for each cancer type based on the Beta-binomial model $\pi_i \sim Beta(\alpha_i, \beta_i)$. | $\alpha_i = \beta_i = 1$ |
| BHM | Bayesian hierarchical model (BHM) assumes a common prior for $\theta_i = logit(\pi_i)$ ; thus, $\theta_i \sim N(\mu, \tau^{-1}), \mu \sim N(\tilde{\mu}, \tilde{\sigma}^2), \tau \sim G(\alpha_\tau, \beta_\tau)$, where $N(\mu, \tau^{-1})$ denotes the normal distribution with mean parameter $\mu$ and variance parameter $\tau^{-1}$. | $\tilde{\mu} = \text{logit}\left(\frac{\pi_{H_0}+\pi_{H_1}}{2}\right), \sigma_0^2 = 10^2$, and $\alpha_\tau = \beta_\tau = 2$. |
| EXNEX | Exchangeability/non-exchangeability model (EXNEX) is an extended model of BHM. This model assumes the two exchangeability models (EX1: $\theta_i^{EX1} \sim N(\mu_{EX1}, \tau_{EX1}^{-1}), \mu_{EX1} \sim N(\tilde{\mu}_{EX1}, \tilde{\sigma}_{EX1}^2), \tau_{EX1} \sim HN(scale = s_\tau)$ ; EX2: $\theta_i^{EX2} \sim N(\mu_{EX2}, \tau_{EX2}^{-1}), \mu_{EX2} \sim N(\tilde{\mu}_{EX2}, \tilde{\sigma}_{EX2}^2), \tau_{EX2} \sim HN(scale = s_\tau)$ ) and non-exchangeability model ( $\theta_i^{NEX} \sim N(\mu_i, \sigma_i^2)$ ), where $HN(scale = s_\tau)$ denotes the half-normal distribution with scale parameter $s_\tau$. The posterior $\pi_i$ is estimated from the three models based on a categorical distribution $C(\delta_i^{EX1}, \delta_i^{EX2}, \delta_i^{NEX})$. | $\tilde{\mu}_{EX1} = logit(\pi_{H_0}), \tilde{\mu}_{EX2} = logit(\pi_{H_1})$, $\tilde{\sigma}_{EX1}^2 = g(\pi_{H_0}), \tilde{\sigma}_{EX2}^2 = g(\pi_{H_1})$, $\mu_i = logit\left\{\frac{\pi_{H_0}+\pi_{H_1}}{2}\right\}, \sigma_i^2 = g\left\{\frac{\pi_{H_0}+\pi_{H_1}}{2}\right\}$, $s_\tau = 0.125$, and $(\delta_i^{EX1}, \delta_i^{EX2}, \delta_i^{NEX}) = (0.33, 0.33, 0.34)$, where $g(\pi) = 1/\pi + 1/(1-\pi) - 1$. |
| MEM | This approach uses the framework of a Bayesian model averaging. Multisource exchangeability model (MEM) assumes $2^{\binom{I}{2}}$ models based on combinations of exchangeability or non-exchangeability in all possible pairs between two cancer types. The posterior response rate for each cancer type is calculated by averaging the posterior response rates in each model using the posterior | $p_m = 0.5$, $\alpha_i = \beta_i = 1$ |
| | probability $p_m$ $\left(m = 1, \cdots, \binom{I}{2}\right)$ for each model plausibility. The prior response rate for cancer type $i$ is assumed as $\pi_i \sim Beta(\alpha_i, \beta_i)$. | |
| BBM-JS | Beta-binomial model with the Jensen–Shannon divergence (BBM-JS) is a BBM where each cancer type borrows the information from the other cancer types with discounting based on similarity $S_{ij}$ using the Jensen–Shannon divergence between cancer types $i$ and $j$. The posterior $\pi_i$ is calculated by $Beta\left(\alpha_i + \sum_{j=1}^{I} S_{ij}^{\epsilon} x_j I(S_{ij} > \tau), \beta_i + \sum_{j=1}^{I} S_{ij}^{\epsilon} (n_j - x_j) I(S_{ij} > \tau)\right)$, where $\epsilon$ is the tuning parameter, $\tau \in (0,1)$ is the cut-off value of $S_{ij}$, and $I(\cdot)$ is the indicator function. | $\alpha_i = \beta_i = 1$, $\epsilon = 2$ and $\tau = 0.5$. |
| BUPD-D | Each cancer type borrows information from other cancer types through $M$ and $w_{ij}$. BUPD-D uses a Dirichlet prior with parameter $\tilde{z}_l$ as a hyper-prior on $\boldsymbol{w}$ based on Equation (6) and uniform prior $Uniform(0, \widetilde{M})$ as a hyper-prior on $M$. | $\tilde{z}_l = 1$ $\left(l = 1, \cdots, \frac{I(I-1)}{2}\right)$, $\widetilde{M} = 72$ |
| BUPD-JS | The same framework used for information borrowing in BUPD-D is used here. BUPD-JS determines $\boldsymbol{w}$ transposed from the Jensen–Shannon divergence in Equation (8) and uses the fixed values of $M = 72$. | $M = 72$ |
| BUPD-JSH | The same framework used for information borrowing in BUPD-D is used here. BUPD-JSH determines $\boldsymbol{w}$ transposed from the Jensen–Shannon divergence in Equation (9) and uses $s \sim G(\tilde{s}, \tilde{s})$ and $M \sim Uniform(0, \widetilde{M})$. | $\tilde{s} = 0.01$, $\widetilde{M} = 72$ |

For each method, the pre-specified cut-off value of $c$ was calibrated to control the cancer-specific type 1 error rate to 0.05 under the scenario with no effective cancer type (i.e., scenario 1). We implemented 2000 simulations for eight scenarios of the true response rate $\pi_i$ (Table 2). The performance indices were cancer-specific power (i.e., the proportion of declaring drug efficacy for each truly effective cancer type among 2000 simulations) and cancer-specific type 1 error rate (i.e., the proportion of declaring drug efficacy for each truly ineffective cancer type among 2000 simulations). The cancer-specific prior effective sample size (i.e., $\alpha_i + \beta_i$ in a Beta prior) was also evaluated for BBM-NB, MEM, BBM-JS, and BUPD-JS (see Section S3 in the Supplemental Material). This evaluation was not included in BHM, EXNEX, BUPD-D, and BUPD-JSH because these methods cannot analytically calculate the effective sample size for the prior of response rate because the informative priors of these methods are not a beta prior. The posterior mean of parameter $s$ in BUPD-JSH and the posterior means of $M$ and $w_{ij}$ among the three proposed methods were investigated. Using the R package "r2jags," we performed 2000 iterations for "burn-in," following which every second sample was retained from 20000 additional iterations, providing a total of 10000 MCMC samples for BHM, EXNEX, BUPD-D, and BUPD-JSH. BBM-NB, MEM, BBM-JS, and BUPD-JS are implemented without the MCMC method. For each simulation, the computational time is approximately 0.01, 0.38, 1.51, 12.7, 0.05, 1.05, 0.01, and 1.26 s in BBM-NB, BHM, EXNEX, MEM, BBM-JS, BUPD-D, BUPD-JS, and BUPD-JSH, respectively. The R codes for BUPD-D, BUPD-JS, and BUPD-JSH were released at https://github.com/rkitabay/UIP_basket.

**Table 2.** True response rates of $\pi_i$ in eight scenarios.

| Scenario | $i = 1$ | 2 | 3 | 4 | 5 | 6 |
|---|---|---|---|---|---|---|
| 1 | 0.10 | 0.10 | 0.10 | 0.10 | 0.10 | 0.10 |
| 2 | 0.10 | 0.10 | 0.10 | 0.10 | 0.10 | **0.40** |
| 3 | 0.10 | 0.10 | 0.10 | 0.10 | **0.40** | **0.40** |
| 4 | 0.10 | 0.10 | 0.10 | **0.40** | **0.40** | **0.40** |
| 5 | 0.10 | 0.10 | **0.40** | **0.40** | **0.40** | **0.40** |
| 6 | 0.10 | **0.40** | **0.40** | **0.40** | **0.40** | **0.40** |
| 7 | **0.40** | **0.40** | **0.40** | **0.40** | **0.40** | **0.40** |
| 8 | 0.05 | 0.10 | **0.20** | **0.30** | **0.40** | **0.50** |

Note: Bold font indicates the effective cancer type.

## 3.2. Comparison of the proposed and existing methods

Figure 1 shows the cancer-specific type 1 error rate and power of the eight methods in scenarios 1–8. In scenario 1, with no effective cancer type, all methods showed a type 1 error rate of 4.3–5.7% for each cancer type. In scenario 2, with only one effective cancer type, the three proposed methods had comparable power with BBM-NB, BHM, and EXNEX, and higher power than MEM and BBM-JS by more than 10%, whereas both the existing and proposed methods maintained type 1 error rates of less than 10%. MEM and BBM-JS diminished the power by allowing the effective cancer type (i.e., cancer type 6) to strongly borrow information from other ineffective cancer types, whereas our proposed methods avoided reducing the power by controlling the strength of information borrowing (i.e., setting the value of $M$ equal to or less than $N$), as detailed in Section S3 of the Supplemental Material.

In Scenario 3 of Figure 1, the type 1 error rates and powers of the three proposed methods are almost the same as those of BHM and EXNEX. The powers of the MEM were the lowest of all the methods in scenario 3. In scenario 4, with three effective and three ineffective cancer types, BUPD-JSH had lower type 1 error rates than BHM, MEM, BBM-JS, BUPD-D, and BUPD-JS by approximately 3–10%, equal power to BHM, BBM-JS, and BUPD-D, and higher power than MEM by approximately 3%. In BUPD-JSH, the posterior mean of the parameter $s$ tended to show lower values (i.e., smaller differences in observed response rates between two cancer types resulted in much higher $w_{ij}$, and larger differences led to much lower $w_{ij}$) under the scenarios with heterogeneous response rates (e.g., scenarios 2–5 and 8), while tended to show higher values (i.e., $w_{12} \approx \cdots \approx w_{I(I-1)} \approx 1/I(I-1)$) under the scenarios with homogeneous response rates (e.g., scenarios 1 and 7), as presented Table S2 in the Supplemental Material. This tendency contributed that BUPD-JSH reduced the type 1 error rates by more effectively controlling the values of $w_{ij}$ using the parameter $s$ and addressing the heterogeneity in response rates compared to BUPD-D and BUPD-JS (see Section S5 in the Supplemental Material).

In scenario 5 of Figure 1, BUPD-D and BUPD-JS had lower type 1 error rates than BHM, MEM, and BBM-JS by approximately 1–11%, and almost the same powers as BHM, MEM, and BBM-JS. BUPD-JSH had lower type 1 error rates than EXNEX by approximately 1%, and higher power than EXNEX by approximately 1%. In Scenario 6, with only one ineffective cancer type, BHM and MEM demonstrated 30% or more type 1 error rates, but BBM-JS, BUPD-D, and BUPD-JS reduced the inflation of type 1 error rates compared to BHM and MEM by 8–14%, while maintaining the same powers of more than 94% as those of BHM and MEM. BBM-JS and our proposed methods addressed the heterogeneity in the response rate for each cancer type compared with BHM and MEM. Among the three proposed methods, the type 1 error rates of BUPD-JS and BUPD-JSH were approximately 9 and 13% lower than those of BUPD-D, respectively. In particular, BUPD-JSH had lower type 1 error rates than EXNEX by 2.8%, and the same power of approximately 91% as EXNEX. In scenario 7, with all effective cancer types, the power of BBM-NB was lower than that of the other methods by more than 10%. BHM and MEM had the highest power among all the methods, followed by BUPD-D and BBM-JS.

In scenario 8, which is a practical setting with all different true response rates, BHM and BUPD-JS demonstrated the highest power for cancer types with intermediate response rates (i.e., cancer types 3 and 4) among all methods. BUPD-JSH had a higher power for cancer types 3 and 4 than EXNEX by 3.2 and 4.5%, respectively, while maintaining the same type 1 error rates for cancer types 1 and 2 at approximately 3 and 13%, respectively, compared to EXNEX. As EXNEX set the NEX model with the mean of $\theta_i$ as the log odds of the medium response rate (i.e., $(\pi_{H_0} + \pi_{H_1})/2$), cancer types 3 and 4 did not borrow the information from the effective cancer types, resulting in diminishing the power. In contrast, BUPD-JSH maintained high power in cancer types 3 and 4 by not requiring the model to be pre-specified based on the response rates. Although the three proposed methods reduced the type 1 error rate in scenarios, including both effective and ineffective cancer types (e.g., scenarios 5 and 6), compared to the existing methods by effectively addressing the heterogeneity in response rates, the three methods showed a high power of more than 90% in scenario 7 by strongly borrowing information achieved through an increase in the value of $M$. The posterior mean of parameter $M$ of the three proposed methods showed higher values in the scenarios under homogeneous response rates (e.g., scenarios 1 and 7) and lower values in the scenario under heterogeneous response rates (e.g., scenarios 2–5 and 8), as presented in Table S3 in the Supplemental Material.

Across the eight scenarios, BBM-NB, BHM, EXNEX, MEM, BBM-JS, BUPD-D, BUPD-JS, and BUPD-JSH had average type 1 error rates of 4.7, 12.2, 8.7, 14.2, 12.7, 11.1, 11.2, and 8.9% and average powers of 79.3, 90.6, 87.3, 88.2, 89.1, 89.8, 90.0, and 87.9%, respectively.

In summary, BUPD-D and BUPD-JS had competitive average type 1 error rates and power with BHM, while avoiding the extreme inflation of the type 1 error rate observed in BHM in scenarios with many effective cancer types (e.g., scenario 6). Compared with MEM and BBM-JS with three and four pre-specified parameters, BUPD-D and BUPD-JS had lower average type 1 error rates of 1.5–3.1%, while they had higher average power of 0.7–1.8%. BUPD-JSH showed almost the same average type 1 error rates and power as EXNEX, including ten pre-specified parameters, but BUPD-JSH had an improved type 1 error rate and power compared to EXNEX in several scenarios (e.g., scenarios 6 and 8).

BUPD-JS and BUPD-JSH demonstrated superior performance in response rate estimation compared to other methods, as measured by the posterior means and the widths of the 95% equal-tailed credential interval for response rates, as detailed in Section S6 of the Supplemental Material.

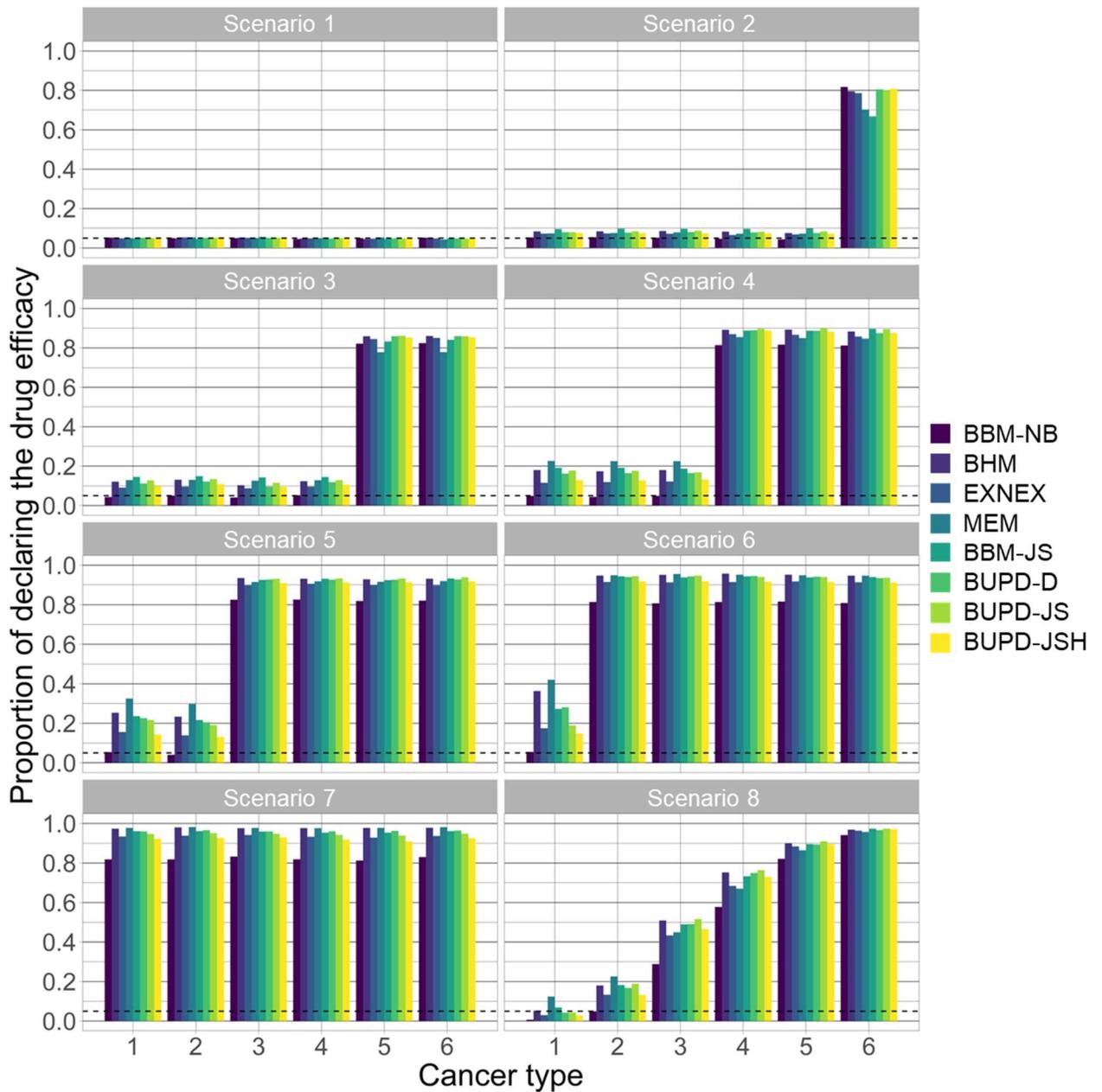

**Figure 1.** Cancer-specific type 1 error rate and power of the eight methods under scenarios 1–8. Dotted line indicates 0.05.

### *3.3. Operating characteristics under different total amount of information*

We also conducted sensitivity analyses to examine the operating characteristics of BUPD-D using different settings of $M$. In the same settings of Section 3.1, the cancer-specific power and type 1 error rate of BUPD-D using $\widetilde{M} = 54$ (BUPD-D-54), $\widetilde{M} = 36$ (BUPD-D-36), and $\widetilde{M} = 18$ (BUPD-D-18) were compared with those of the original method using $\widetilde{M} = 72$ (BUPD-D-72).

Figure 2 shows the cancer-specific type 1 error rate and power of BUPD-D-72, BUPD-D-54, BUPD-D-36, and BUPD-D-18 under scenarios 1–8. The power of BUPD-D-72 was the highest among the four methods in scenarios 3–8, followed by those of BUPD-D-54, BUPD-D-36, and BUPD-D-18. The type 1 error rate of BUPD-D-72 was also the highest among the four methods in scenarios 2–6 and 8, followed by those of BUPD-D-54, BUPD-D-36, and BUPD-D-18. Thus, increasing $\widetilde{M}$ increases both the power and type 1 error rates, while decreasing $\widetilde{M}$ decreases both the power and type 1 error rates. BUPD-D can be modified to achieve the desired power (and type 1 error rate) by changing the value of the parameter $\widetilde{M}$. Similar operating characteristics using different settings of $M$ were also found in BUPD-JS (data not shown). Across the eight scenarios, BUPD-D-18, BUPD-D-36, BUPD-D-54, and BUPD-D-72 had average type 1 error rates of 5.1, 7.4, 9.4, and 11.1%, and average powers of 80.8, 85.7, 88.4, and 89.8%, respectively. Compared to BBM-NB in Section 3.2, BUPD-D-18 had almost the same average type 1 error rate while improving the average power by 1.5%. The reason for the similar average type 1 error rate between BUPD-D-18 and BBM-NB is that BUPD-D-18 borrows the information of only 18 patients (i.e., each cancer type borrows the information of a total of three patients from the other cancer types on average) in the basket trial. Furthermore, as BUPD-D addresses heterogeneity in the response rate for each cancer type, BUPD-D-18 improves the power by borrowing information between effective cancer types compared with BBM-NB in each scenario.

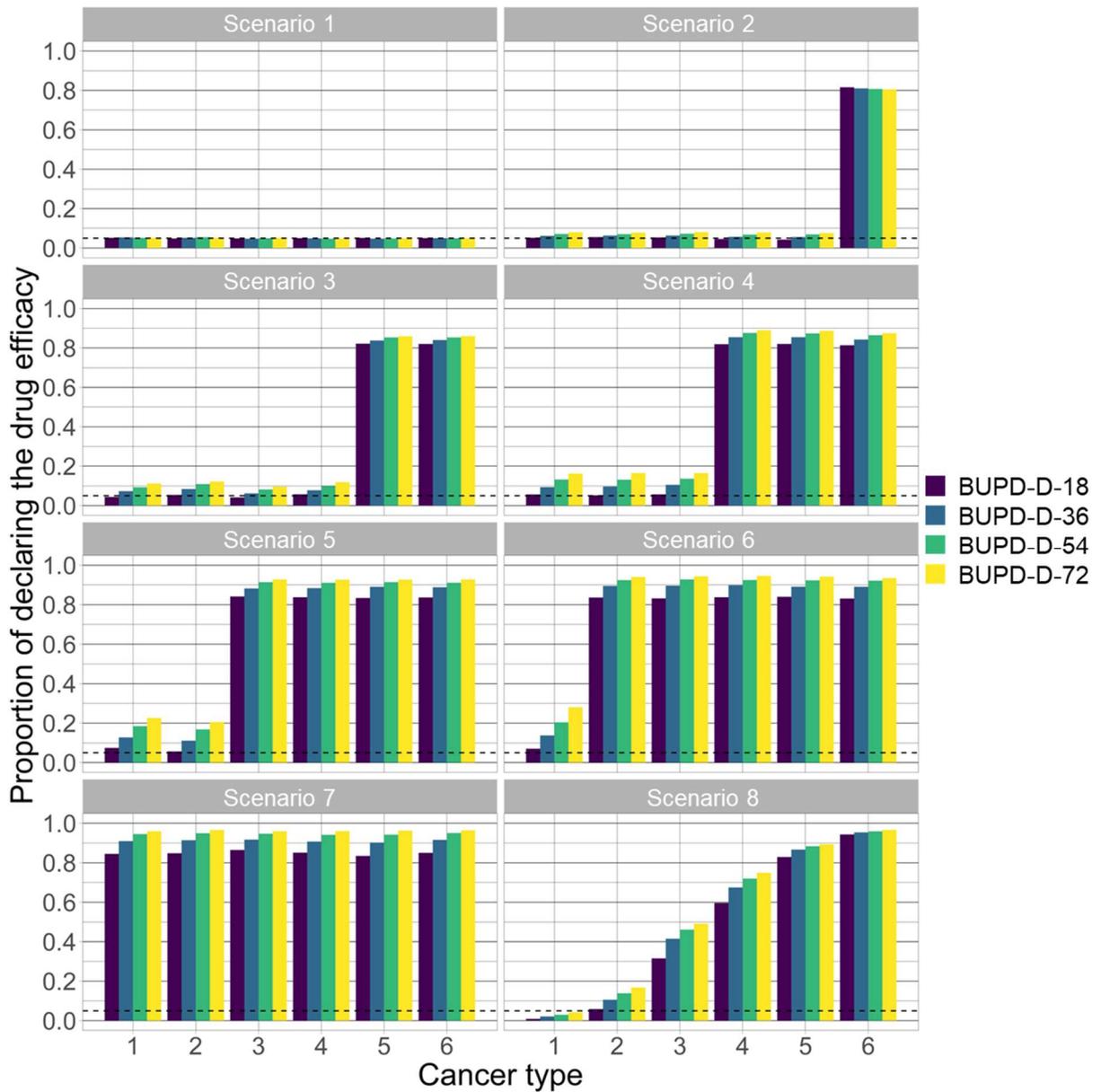

**Figure 2.** Cancer-specific type 1 error rate and power of BUPD-D-72, BUPD-D-54, BUPD-D-36, and BUPD-D-18 under scenarios 1–8. Dotted line indicates 0.05.

## 4. Application of the proposed method to real data

This section presents the application of the proposed methods using real data from a basket trial on vemurafenib,[4] in which a response rate of 15% was considered low and 35% was considered desirable and indicative of efficacy. This trial included six cohorts: non-small-cell lung cancer (NSCLC), colorectal cancer with vemurafenib (CRC-V), colorectal cancer with vemurafenib and cetuximab (CRC-VC), cholangiocarcinoma (CCA), Erdheim–Chester disease/Langerhan's cell histiocytosis (ECD/LCH), and anaplastic thyroid cancer (ATC). Eighty-four eligible patients were enrolled in the six cohorts. We compared the eight methods discussed in Section 3.1, assuming $\pi_{H_0} = 0.15, \pi_{H_1} = 0.35,$ and $I = 6$. We set $\widetilde{M} = 84$ (i.e., total number of patients in the six cohorts) for the three proposed methods, and the other parameters for each method were the same as those used in the simulation studies described in Section 3.

Table 3 shows the posterior mean of the response rates, 95% equal-tailed interval (ETI), and posterior probability (PP) of $\Pr(\pi_i > \pi_{H_0}|\boldsymbol{n}, \boldsymbol{x})$ for each method. For all methods, the posterior means of response rates for CRC-V and CRC-VC were lower than the null response rate of 15%, whereas those for NSCLC and ECD/LCH were higher than the alternative response rate of 35%. BUPD-JS, BUPD-JSH, and EXNEX exhibited lower posterior mean response rates than BUPD-D and BHM for CRC-V and CRC-VC (presumably, the therapeutic effect of the drug on these cancer types was low) and higher posterior mean response rates than BUPD-D and BHM for NSCLC and ECD/LCH (presumably, the therapeutic effect of the drug on these cancer types was high). The widths of ETI for each cancer type in MEM and BBM-JS tended to be smaller than those of the other methods (e.g., for ECD/LCH, the widths of ETI in NB, BHM, EXNEX, MEM, BBM-JS, BUPD-D, BUPD-JS, and BUPD-JSH were 46.4, 44.8, 39.2, 31.2, 28.6, 40.9, 33.8, and 36.8, respectively), whereas those of BUPD-D and BUPD-JSH were comparable, with values approximately 40% similar to those of EXNEX. BUPD-D and BUPD-JSH avoided extremely strong information borrowing by considering heterogeneity in the response rates in the trial, resulting in a large variance in the posterior distribution (i.e., wide ETI). For NSCLC and ECD/LCH, the PPs were close to 100% for all methods, which is consistent with the conclusion of Hyman et al. (2015) that the two cohorts were effective. However, for ATC, the PP varied widely depending on the method (e.g., minimum value was 74.7% in BHM and maximum value was 100% in BBM-JS). Specifically, PP of ATC was > 90% in MEM, BBM-JS, BUPD-JS, and BUPD-JSH, whereas the original trial, which used the adaptive Simon two-stage design[27] to evaluate the efficacy for each cohort, was unable to declare the efficacy of drug for the cancer type. These methods would have assisted the investigators of the original trials in declaring the drug efficacy against ATC.

**Table 3.** Posterior means of the response rate $\hat{\pi}_i$, 95% equal-tailed intervals (ETIs), and posterior probabilities (PPs) of $Pr(\pi_i > \pi_{H_0} | n, x)$ estimated for six cohorts using eight existing and proposed methods.

| Cohort $x/n$ (%) | | BBM-NB | BHM | EXNEX | MEM | BBM-JS | BUPD-D | BUPD-JS | BUPD-JSH |
|---|---|---|---|---|---|---|---|---|---|
| NSCLC 8/19 (42.1) | $\hat{\pi}_i$ | 42.9 | 37.6 | 39.1 | 40.6 | 41.6 | 36.3 | 39.0 | 39.8 |
| | ETI | (23.1–63.9) | (18.9–58.4) | (22.2–57.7) | (25.7–56.4) | (27.7–56.3) | (19.0–56.5) | (23.6–55.5) | (23.4–57.4) |
| | PP | 99.9 | 99.4 | 99.8 | 100.0 | 100.0 | 99.5 | 100.0 | 99.9 |
| CRC-V 0/10 (0.0) | $\hat{\pi}_i$ | 8.3 | 8.4 | 3.9 | 7.0 | 9.5 | 9.0 | 5.4 | 4.9 |
| | ETI | (0.2–28.5) | (0.5–25.3) | (0.0–22.7) | (1.6–16.1) | (2.8–19.5) | (0.1–27.0) | (0.3–16.8) | (0.0–18.7) |
| | PP | 16.7 | 15.1 | 5.6 | 4.7 | 10.9 | 19.0 | 4.0 | 5.6 |
| CRC-VC 1/26 (3.8) | $\hat{\pi}_i$ | 7.1 | 7.3 | 5.4 | 6.6 | 9.1 | 7.5 | 4.5 | 5.4 |
| | ETI | (0.9–19.0) | (1.1–18.4) | (0.3–47.6) | (1.4–15.5) | (2.7–19.0) | (0.9–19.1) | (0.3–13.6) | (0.5–16.3) |
| | PP | 7.2 | 6.4 | 4.6 | 3.4 | 9.6 | 7.7 | 1.5 | 3.6 |
| CCA 1/8 (12.5) | $\hat{\pi}_i$ | 20.0 | 15.2 | 17.0 | 11.2 | 14.9 | 17.3 | 17.1 | 16.5 |
| | ETI | (2.8–48.2) | (2.1–39.4) | (1.1–45.7) | (4.2–21.5) | (5.9–26.9) | (2.8–39.3) | (6.6–31.4) | (3.2–35.4) |
| | PP | 59.9 | 42.6 | 44.3 | 20.7 | 45.0 | 54.3 | 59.2 | 52.6 |
| ECD/LCH 6/14 (42.9) | $\hat{\pi}_i$ | 43.8 | 36.7 | 38.9 | 40.5 | 41.6 | 35.9 | 38.2 | 39.3 |
| | ETI | (21.3–67.7) | (16.1–60.9) | (19.8–59.0) | (25.4–56.6) | (27.6–56.2) | (17.4–58.3) | (22.0–55.8) | (21.5–58.3) |
| | PP | 99.6 | 98.2 | 99.4 | 100.0 | 100.0 | 98.9 | 99.9 | 99.7 |
| ATC 2/7 (28.6) | $\hat{\pi}_i$ | 33.3 | 24.8 | 31.7 | 33.9 | 37.5 | 26.0 | 30.0 | 29.6 |
| | ETI | (8.5–65.1) | (5.7–54.3) | (5.4–55.1) | (20.2–49.4) | (24.3–51.6) | (7.4–51.3) | (15.5–47.0) | (11.4–52.2) |
| | PP | 89.5 | 74.7 | 84.7 | 91.1 | 100.0 | 83.3 | 97.9 | 93.5 |

Table 4 shows the posterior means of $Mw_{ij}$ for BUPD-D, BUPD-JS, and BUPD-JSH for the six cohorts. These values indicate how many patients with information mutually borrowed between two cancer types (e.g., NSCLC and ECD/LCH mutually borrowed information of 2.8, 8.2, and 8.4 patients in BUPD-D, BUPD-JS, and BUPD-JSH, respectively). In common with the three proposed methods, a smaller (or larger) difference in the observed response rates between the two cancer types leads to stronger (or weaker) mutual information borrowing. For example, the values of $Mw_{ij}$ between CRC-V and CRC-VC (i.e., the observed response rate difference was 3.8%) were 2.2, 7.3, and 5.6 in BUPD-D, BUPD-JS, and BUPD-JSH, respectively, and the values of $Mw_{ij}$ between CRC-V and ECD/LCH (i.e., the observed response rate difference was 42.9%) were 0.8, 0.1, and 0.2, respectively.

Compared to BUPD-JS and BUPD-JSH, the values of $Mw_{ij}$ for BUPD-D did not vary (e.g., the minimum and maximum values of $Mw_{ij}$ for BUPD-D, BUPD-JS, and BUPD-JSH were $(0.8, 2.8), (0.0, 8.2), (0.1, 8.4)$, respectively), indicating that BUPD-D tended to borrow similar amounts of information between the two cancer types in all combinations. Furthermore, BUPD-D evaluated the response rates for each cancer type as more homogeneous than BUPD-JS and BUPD-JSH (e.g., the posterior means of response rates for almost all cancer types in BUPD-D were close to the overall observed mean of $18/84 \approx 21.4\%$ compared to BUPD-JS and BUPD-JSH in Table 3). We believe that this tendency of BUPD-D led to an increase in the type 1 error rate compared to BUPD-JS and BUPD-JSH (e.g., scenario 6 of Figure 1 in Section 3.2). In comparison between BUPD-JS and BUPD-JSH, as BUPD-JS used the fixed value of $M = 84$ (i.e., the maximum value of the range of $M \sim U(0, 84)$ in BUPD-JSH), the value of $M$ in BUPD-JS was higher than in BUPD-JSH, resulting in that the values of $Mw_{ij}$ of BUPD-JS were higher than those of BUPD-JSH in almost all combinations between cancer types $i$ and $j$. However, the value of $Mw_{ij}$ between NSCLC and ECD/LCH in BUPD-JSH was slightly higher than those in BUPD-JS (i.e., 8.4 vs. 8.2). Because the observed response rates of NSCLC (42.1%) and ECD/LCH (42.9%) were almost the same and both cancer types had many patients (i.e., 19 and 14), it is preferable to assign larger values of $Mw_{ij}$ between the two cancer types than those between the other two cancer types. Thus, BUPD-JSH effectively estimates the value of $w_{ij}$. As the difference in the estimation method for $w_{ij}$ between BUPD-JS and BUPD-JSH lies in whether parameter $s$ is incorporated, this parameter improves the evaluation of heterogeneity in response rates for each cancer type.

**Table 4.** Posterior means of $Mw_{ij}$ for the three proposed methods in six cohorts.

**BUPD-D**

| $Mw_{ij} =$ | CRC-V | CRC-VC | CCA | ECD/LCH | ATC |
|---|---|---|---|---|---|
| NSCLC | 0.9 | 0.9 | 1.3 | 2.8 | 1.8 |
| CRC-V | | 2.2 | 1.6 | 0.8 | 1.2 |
| CRC-VC | | | 1.8 | 0.8 | 1.1 |
| CCA | | | | 1.2 | 1.4 |
| ECD/LCH | | | | | 1.7 |

**BUPD-JS**

| $Mw_{ij} =$ | CRC-V | CRC-VC | CCA | ECD/LCH | ATC |
|---|---|---|---|---|---|
| NSCLC | 0.1 | 0.0 | 1.2 | 8.2 | 5.1 |
| CRC-V | | 7.3 | 3.9 | 0.1 | 0.8 |
| CRC-VC | | | 2.7 | 0.0 | 0.4 |
| CCA | | | | 1.3 | 5.2 |
| ECD/LCH | | | | | 5.6 |

**BUPD-JSH**

| $Mw_{ij} =$ | CRC-V | CRC-VC | CCA | ECD/LCH | ATC |
|---|---|---|---|---|---|
| NSCLC | 0.2 | 0.1 | 0.7 | 8.4 | 3.0 |
| CRC-V | | 5.6 | 2.1 | 0.2 | 0.5 |
| CRC-VC | | | 1.4 | 0.1 | 0.3 |
| CCA | | | | 0.7 | 3.1 |
| ECD/LCH | | | | | 3.5 |

## 5. Discussion

In this study, we have developed a novel Bayesian under-parameterized design to estimate response rates based on a unit information prior to oncology basket trials. In contrast to existing methods, which involve many pre-specified parameters to account for the heterogeneity of response rates among cancer types, BUPD borrows information on response rates through only one (or two) pre-specified parameters. We have proposed three distinct methods for specifying the prior distribution of these parameters, each tailored to address different practical considerations and enhance flexibility in real-world applications. BUPD flexibly controls the type 1 error rate and power by adjusting the parameter that explicitly specifies the strength of the borrowing interpretable as the sample size.

A simulation study revealed that the BUPD (i.e., BUPD-D, BUPD-JS, and BUPD-JSH). BUPD improved the power by approximately 10–14% compared with MEM and BBM-JS under scenarios with only one effective cancer type while maintaining a type 1 error rate of less than 10%, as BUPD controlled the strength of information borrowing to avoid extremely strong borrowing information between effective and ineffective cancer types. BUPD-D and BUPD-JS reduced the cancer-specific type 1 error rate by approximately 3–17% compared with BHM under scenarios with several effective cancer types (e.g., scenarios 5–6), while maintaining a similar cancer-specific power of 93–95%. BUPD-JSH improved the power by approximately 4% compared with EXNEX under practical scenarios with truly different response rates for each cancer type (e.g., Scenario 8), while maintaining a similar type 1 error rate of less than 13%. Comparing with BHM with only one parameter (i.e., the variance of $\theta_i$) to address heterogeneity, BUPD successfully avoided borrowing the information between effective and ineffective cancer types by introducing the weight parameter for each pairwise combination between two cancer types, resulting in reducing the inflation of the type 1 error rate in scenarios with few ineffective cancer types.

Here, simulation studies and sensitivity analyses revealed that the respective differences of the average power and type 1 error rate between BUPD-D and BUPD-JS2, respectively; average type 1 error rates of 5.1 and 11.1%, and average powers of 80.8 and 90.0% in BUPD-D-18 and BUPD-D-72, respectively). Thus, the prior specification for $M$ is more important than that for $w_{ij}$ to control the average type 1 error rate or power. In early-phase trials, such as exploratory trials, where the true response rates are usually unknown, the parameter $M$ or $\widetilde{M}$ of its hyper-prior in BUPD should be set to a larger value (e.g., total number of patients enrolled in the basket trial) to improve the power while allowing for the inflation of type 1 error rate at a certain level. Conversely, in trials in which the control of type 1 error rates is crucial, $M$ or $\widetilde{M}$ should be set to a smaller value (e.g., half or a quarter of the total number of patients enrolled in the basket trial). Given that the value of $M$ can be understood by non-statisticians as the sample size, we recommend consulting with the investigators to determine whether the chosen value for this parameter is appropriate for the trial.

Across the eight scenarios in our simulation studies, BUPD-JSH showed favorable performance with average type 1 error rates approximately 2% lower than those observed in BUPD-D and BUPD-JS (average type 1 error rate was approximately 11% for both BUPD-D and BUPD-JS). Furthermore, BUPD-JSH reduced the type 1 error rate by approximately 3–4% compared with BUPD-D and BUPD-JS in scenario 4 (Table 2) while maintaining a similar power of 90% by effectively controlling the values of weight parameters and parameter $M$. Similar to BUPD-JS, BUPD-JSH was also found to have reduced the average bias of posterior means of response rates by approximately 2% when compared with BUPD-D while reducing the average width of the 95% equal-tailed credential intervals by 1.3–3.7%. Although BUPD-JSH is the first choice in terms of the preferred operating characteristics, BUPD-JS can be performed without the MCMC method, thereby significantly reducing the computational burden. This advantage makes it particularly well-suited for large basket trials (e.g., those involving more than ten cancer types), where evaluating operating characteristics across numerous scenarios is essential. BUPD-D could also be an alternative based on the simple structure of the non-informative hyper-prior on $w_{ij}$ when the expected difference in operating characteristics compared to BUPD-JSH is small (e.g., in basket trials with fewer cancer types and limited information borrowing).

Our simulation study revealed that BHM had the highest average power of approximately 91% across both the existing and proposed methods; thus, BHM would be an effective method if the type 1 error rate inflation is acceptable. The MEM shows the same power as the BHM under scenarios with all effective cancer types, positioning this method as an effective alternative in such cases. Notably, MEM requires substantial computational resources for the implementation in the trials with many cancer types (e.g., MEM necessitates the calculation of $2^{\binom{8}{2}} \approx 2.7 \times 10^8$ models in trials with eight cancer types). When heterogeneous response rates were expected among cancer types, BBM-JS, EXNEX, and the three proposed methods demonstrated preferable operating characteristics in terms of less inflation of the type 1 error rate compared to BHM and MEM (e.g., scenario 6). Moreover, similar to BUPD-JS, BBM-JS can estimate posterior response rates without using the MCMC method. However, caution is advised when using BBM-JS, especially in trials where the majority of cancer types may be ineffective, as it may lead to a reduction in power, and BBM-JS requires careful consideration for the setting of the two tuning parameters. If minimizing the type 1 error rate inflation is a priority, EXNEX emerges as an effective method because it has the lowest average type 1 error rate among the compared methods, whereas the three proposed methods using a small value of $M$ can further reduce the type 1 error rate compared to EXNEX. Note that the EXNEX requires an appropriate model structure, including the null and alternative response rates. For strict control of the type 1 error rate, BBM-NB remains a reasonable choice.

To account for the heterogeneity of response rates among cancer types, we used the Dirichlet prior and Jensen–Shannon divergence for parameter $w_{ij}$; however, other approaches or measures (e.g., Hellinger divergence and two-sided p-value of Fisher's exact test) are also available. Furthermore, the cut-off value of $w_{ij}$ can be incorporated to truncate information borrowing by replacing the value of $w_{ij}$ with zero if the response rates of the two cancer types are obviously dissimilar. In this study, we assumed a basket trial with a binary endpoint. Recent studies have also proposed information borrowing methods for continuous endpoints with a normal distribution.[23,28] Our BUPD framework can be easily extended for continuous endpoints.

**Funding**

This work was partially supported by JSPS KAKENHI (grant number: 24K02906) (Grant in-Aid for Scientific Research B) and Japan Agency for Medical Research and Development (grant number: JP24mk0121268).

**Declaration of Conflicting Interests**

The authors declare that there is no conflict of interest.

**Data availability statement**

This study did not involve the generation or analysis of datasets, thus data sharing is not applicable.

**Supplemental material**

Supplemental material for this study is provided online.

**References**

1. Heinrich MC, Joensuu H, Demetri GD, et al. Phase II, open-label study evaluating the activity of imatinib in treating life-threatening malignancies known to be associated with imatinib-sensitive tyrosine kinases. *Clin Cancer Res.* 2008;14**:**2717–2725.

2. Kaufman B, Shapira-Frommer R, Schmutzler RK, et al. Olaparib monotherapy in patients with advanced cancer and a germline BRCA1/2 mutation. *J Clin Oncol.* 2015;33:244–250.
3. Hyman DM, Piha-Paul SA, Won H, et al. HER kinase inhibition in patients with HER2- and HER3-mutant cancers. *Nature.* 2018;554:189–194.
4. Hyman DM, Puzanov I, Subbiah V, et al. Vemurafenib in multiple nonmelanoma cancers with BRAF V600 mutations. *N Engl J Med.* 2015;373:726–736.
5. Thall PF, Wathen JK, Bekele BN, et al. Hierarchical Bayesian approaches to phase II trials in diseases with multiple subtypes. *Stat Med.* 2003;22:763–780.
6. Berry SM, Broglio KR, Groshen S, et al. Bayesian hierarchical modeling of patient subpopulations: efficient designs of phase II oncology clinical trials. *Clin Trials.* 2013;10:720–734.
7. Freidlin B and Korn EL. Borrowing information across subgroups in phase II trials: Is it usedul? *Clin Cancer Res.* 2013;19:1326–1334.
8. Neuenschwander B, Wandel S, Roychoudhury S, et al. Robust exchangeability designs for early phase clinical trials with multiple strata. *Pharm Stat.* 2015;15:123–134.
9. Chen C and Hsiao CF. Bayesian hierarchical models for adaptive basket trial design. *Pharm Stat.* 2023;22:531–546.
10. Daniells L, Mozgunov P, Bedding A, et al. A comparison of Bayesian information borrowing methods in basket trials and a novel proposal of modified exchangeability-nonexchangeability method. *Stat Med.* 2023;42:4392–4417.
11. Chu Y and Yuan Y. A Bayesian basket trial design using a calibrated Bayesian hierarchical model. *Clin Trials.* 2018;15:149–158.
12. Jin J, Riviere MK, Luo X, et al. Bayesian methods for the analysis of early-phase oncology basket trials with information borrowing across cancer types. *Stat Med.* 2020;39:3459–3475.
13. Chen N and Lee JJ. Bayesian hierarchical classification and information sharing for clinical trials with subgroups and binary outcomes. *Biom J.* 2018;61:1219–1231.
14. Chen N and Lee JJ. Bayesian cluster hierarchical model for subgroup borrowing in the design and analysis of basket trials with binary endpoints. *Stat Methods Med Res.* 2020;29:2717–2732.
15. Jiang L, Nie L, Yan F, et al. Optimal Bayesian hierarchical model to accelerate the development of tissue-agnostic drugs and basket trials. *Contemp Clin Trials.* 2021;107:106460.
16. Lyu J, Zhou T, Yuan S, et al. MUCE: Bayesian hierarchical modeling for the design and analysis of phase 1b multiple expansion cohort trials. *J R Stat Soc Ser C Appl Stat.* 2023;72:649–669.
17. Zhou T and Ji Y. RoBoT: A robust Bayesian hypothesis testing method for basket trials. *Biostatistics.* 2021;22:897–912.

18. Broglio KR, Zhang F, Yu B, et al. A comparison of different approaches to Bayesian hierarchical models in a basket trial to evaluate the benefits of increasing complexity. *Stat Biopharm Res.* 2022;14:324–333.
19. Hobbs BP and Landin R. Bayesian basket trial design with exchangeability monitoring. *Stat Med.* 2018;37:3557–3572.
20. Psioda MA, Xu J, Jiang Q, et al. Bayesian adaptive basket trial design using model averaging. *Biostatistics.* 2019;22:19–34.
21. Fujikawa K, Teramukai S, Yokota I, et al. A Bayesian basket trial design that borrows information across strata based on the similarity between the posterior distributions of the response probability. *Biom J.* 2020;62:330–338.
22. Jin H and Yin G. Unit information prior for adaptive information borrowing from multiple historical datasets. *Stat Med.* 2021;40:5657–5672.
23. Zheng H and Wason J. Borrowing of information across patient subgroups in a basket trial based on distributional discrepancy. *Biostatistics.* 2022;23:120–135.
24. Cunanan KM, Iasonos A, Shen R, et al. Variance prior specification for a basket trial design using Bayesian hierarchical modeling. *Clin Trials.* 2019;16(2):142–153.
25. Jiang L, Nie L, Yan F, et al. Optimal Bayesian hierarchical model to accelerate the development of tissue-agnostic drugs and basket trials. *Contemp Clin Trials.* 2021;107:106460.
26. Chen C and Hsiao CF. Bayesian hierarchical models for adaptive basket trial designs. *Pharm Stat.* 2023;22(3):531-546.
27. Lin Y and Shih WJ. Adaptive two-stage designs for single-arm phase IIA cancer clinical trials. *Biometrics.* 2004;60:482–490.
28. Ouma LO, Grayling MJ, Wason J, et al. Bayesian modelling strategies for borrowing of information in randomised basket trials. *J R Stat Soc Ser C Appl Stat.* 2022;71:2014–2037.

**Supplemental material**
**BUPD: A Bayesian under-parameterized basket design with the unit information prior in oncology trials**

## S1. Property of parameter $s$ in BUPD-JSH

BUPD-JSH introduces the parameter $s$ when transforming the Jensen–Shannon divergence $d_{ij}$ into $w_{ij}$ in Equation (9). Let $d_{i^*j^*}$ and $w_{i^*j^*}$ denote the minimum $d_{ij}$ among all combinations between two cancer types (i.e., $I(I-1)/2$ combinations) and the corresponding weight transformed from $d_{i^*j^*}$, respectively, and $d_{(-i^*j^*)}$ and $w_{(-i^*j^*)}$ denote the arbitrary $d_{ij}$ and $w_{ij}$ of the combination between cancer types $i$ and $j$ other than cancer types $i^*$ and $j^*$. Assuming $s > 0$ and $d_{ij} > 0$, the parameter $s$ has the following two properties to control the values of $w_{ij}$:

Property 1: Under $s \to \infty$, all values of $w_{ij}$ are equal, i.e., $w_{12} = \cdots = w_{(I-1)I} = 1/\{I(I-1)\}$

Property 2: Under $s \to 0$, $w_{i^*j^*} = w_{j^*i^*} = 0.5$, and $w_{(-i^*j^*)} = 0$

Proof 1: For the numerator in Equation (9), when $s \to \infty$, we obtain $d_{ij}/s \to 0$ and $\exp(-d_{ij}/s) \to 1$. For the denominator, $\sum_{i=1}^{I} \sum_{j \in I_{(-i)}} \exp(-d_{ij}/s) \to \sum_{i=1}^{I} \sum_{j \in I_{(-i)}} 1 = I(I-1)$. Therefore, the value of $w_{ij}$ between cancer types $i$ and $j$ is $1/I(I-1)$ when $s \to \infty$.

Proof 2: First, we consider $\exp\left(-\frac{d_{(-i^*j^*)}}{s}\right) / \exp\left(-\frac{d_{i^*j^*}}{s}\right)$ under $s \to 0$. As $d_{(-i^*j^*)} - d_{i^*j^*} > 0$, we obtain $\frac{d_{(-i^*j^*)} - d_{i^*j^*}}{s} \to \infty$ based on $s > 0$; therefore,

$$\frac{\exp\left(-\frac{d_{(-i^*j^*)}}{s}\right)}{\exp\left(-\frac{d_{i^*j^*}}{s}\right)} = \exp\left\{\frac{-(d_{(-i^*j^*)} - d_{i^*j^*})}{s}\right\} \to \exp(-\infty) = 0. \quad \text{(S1)}$$

As $\exp(-d_{ij}/s) > 0$ in the arbitrary cancer types $i$ and $j$, we obtain

$$\exp\left(-\frac{d_{i^*j^*}}{s}\right) < \sum_{i=1}^{I} \sum_{j \in I_{(-i)}} \exp\left(-\frac{d_{ij}}{s}\right)$$

$$\frac{1}{\exp\left(-\frac{d_{i^*j^*}}{s}\right)} > \frac{1}{\sum_{i=1}^{I} \sum_{j \in I_{(-i)}} \exp\left(-\frac{d_{ij}}{s}\right)}$$

$$\frac{\exp\left(-\frac{d_{(-i^*j^*)}}{s}\right)}{\exp\left(-\frac{d_{i^*j^*}}{s}\right)} > \frac{\exp\left(-\frac{d_{(-i^*j^*)}}{s}\right)}{\sum_{i=1}^{I} \sum_{j \in I_{(-i)}} \exp\left(-\frac{d_{ij}}{s}\right)}. \quad \text{(S2)}$$

The right side of the last line of Equation (S2) is $w_{(-i^*j^*)}$ in Equation (9), and the left side is close to 0 based on Equation (S1) when $s \to 0$. Therefore, $w_{(-i^*j^*)} \to 0$ when $s \to 0$. As we constrain $w_{ij} = w_{ji}$ and $\sum_{i=1}^{I} \sum_{j \in I_{(-i)}} w_{ij} = 1$ in Section 2.1, $w_{i^*j^*} = w_{j^*i^*} = 0.5$ when $s \to 0$.

## S2. Diagrams of the model structures for BUPD-D, BUPD-JS, and BUPD-JSH

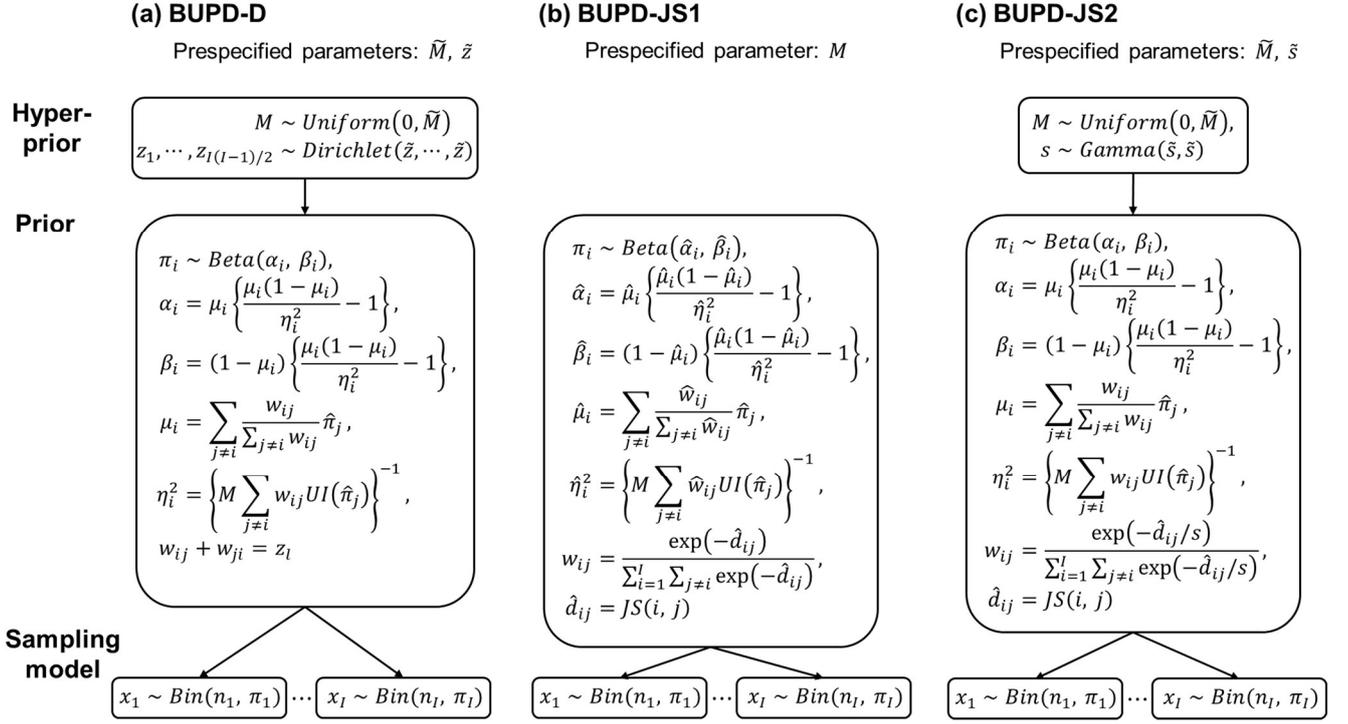

**Figure S1.** Diagrams of the model structures for (a) BUPD-D, (b) BUPD-JS, and (c) BUPD-JSH.

### S3. Cancer-specific prior effective sample size of BBM-NB, MEM, BBM-JS, and BUPD-JS in the simulation study

Table S1 shows the cancer-specific prior effective sample size (i.e., $\alpha_i + \beta_i$ in a Beta prior) of BBM-NB, MEM, BBM-JS, and BUPD-JS under scenarios 1–8 in the simulation study of Section 3. The effective sample size of BBM-NB was 2 in all scenarios because $\alpha_i = \beta_i = 1$ in BBM-NB, and there is no information borrowing. The MEM and BBM-JS showed higher effective sample sizes than BUPD-JS in all scenarios, indicating that the prior distribution for each cancer type in MEM and BBM-JS incorporated more information derived from other cancer types than in BUPD-JS. In other words, each cancer type borrowed information more strongly from the other cancer types in MEM and BBM-JS than in BUPD-JS. In particular, the effective sample sizes of MEM and BBM-JS were more than 12 (i.e., the average number of patients enrolled for each cancer type) in cancer type 6 in scenario 2; thus, the posterior distribution of cancer type 6 included more information from other ineffective cancer types than its own sample size, resulting in reduced power (see cancer type 6 in scenario 2 of Figure 1). In contrast, the effective sample size of BUPD-JS was only 5.4 in cancer type 6. This indicates that BUPD-JS avoids reducing power by not prespecifying the value of $M$ to be extremely high and controlling the strength of information borrowing, unlike MEM and BBM-JS, which lack a parameter corresponding to $M$. In Scenario 7, the effective sample size for each cancer type was approximately 12 in BUPD-JS, and the total effective sample size was 71.3, which was similar to the value of $M$ $(= 72)$.

**Table S1.** Cancer-specific and total prior effective sample size of BBM-NB, MEM, BBM-JS, and BUPD-JS under scenarios 1–8.

| Scenario | Method | $i=1$ | 2 | 3 | 4 | 5 | 6 | Total |
|---|---|---|---|---|---|---|---|---|
| 1 | $\pi_i=$ | 0.10 | 0.10 | 0.10 | 0.10 | 0.10 | 0.10 | |
| | BBM-NB | 2.0 | 2.0 | 2.0 | 2.0 | 2.0 | 2.0 | 12.0 |
| | MEM | 53.6 | 53.6 | 53.7 | 53.6 | 53.5 | 53.7 | 321.9 |
| | BBM-JS | 53.0 | 53.2 | 53.4 | 53.1 | 52.9 | 53.1 | 318.6 |
| | BUPD-JS | 11.2 | 11.2 | 11.2 | 11.2 | 11.1 | 11.1 | 67.1 |
| 2 | $\pi_i=$ | 0.10 | 0.10 | 0.10 | 0.10 | 0.10 | 0.40 | |
| | BBM-NB | 2.0 | 2.0 | 2.0 | 2.0 | 2.0 | 2.0 | 12.0 |
| | MEM | 46.6 | 46.5 | 46.6 | 46.5 | 46.5 | 21.9 | 254.5 |
| | BBM-JS | 46.3 | 46.2 | 46.5 | 46.2 | 46.3 | 19.5 | 251.0 |
| | BUPD-JS | 13.3 | 13.4 | 13.5 | 13.3 | 13.4 | 5.4 | 72.3 |
| 3 | $\pi_i=$ | 0.10 | 0.10 | 0.10 | 0.10 | 0.40 | 0.40 | |
| | BBM-NB | 2.0 | 2.0 | 2.0 | 2.0 | 2.0 | 2.0 | 12.0 |
| | MEM | 39.7 | 39.7 | 39.9 | 39.7 | 24.8 | 24.4 | 208.3 |
| | BBM-JS | 39.6 | 39.5 | 40.0 | 39.4 | 25.1 | 24.8 | 208.4 |
| | BUPD-JS | 14.7 | 14.7 | 14.8 | 14.7 | 9.7 | 9.5 | 78.1 |
| 4 | $\pi_i=$ | 0.10 | 0.10 | 0.10 | 0.40 | 0.40 | 0.40 | |
| | BBM-NB | 2.0 | 2.0 | 2.0 | 2.0 | 2.0 | 2.0 | 12.0 |
| | MEM | 34.2 | 33.9 | 34.0 | 29.0 | 29.4 | 29.3 | 189.8 |
| | BBM-JS | 33.6 | 33.2 | 33.4 | 30.9 | 31.4 | 31.4 | 193.9 |
| | BUPD-JS | 14.0 | 13.7 | 13.8 | 13.1 | 13.2 | 13.2 | 80.9 |
| 5 | $\pi_i=$ | 0.10 | 0.10 | 0.40 | 0.40 | 0.40 | 0.40 | |
| | BBM-NB | 2.0 | 2.0 | 2.0 | 2.0 | 2.0 | 2.0 | 12.0 |
| | MEM | 28.5 | 27.9 | 33.7 | 33.7 | 33.5 | 33.7 | 190.9 |
| | BBM-JS | 26.8 | 25.9 | 37.0 | 37.1 | 36.6 | 37.0 | 200.4 |
| | BUPD-JS | 10.7 | 10.4 | 14.7 | 14.8 | 14.6 | 14.8 | 80.0 |
| 6 | $\pi_i=$ | 0.10 | 0.40 | 0.40 | 0.40 | 0.40 | 0.40 | |
| | BBM-NB | 2.0 | 2.0 | 2.0 | 2.0 | 2.0 | 2.0 | 12.0 |
| | MEM | 24.6 | 38.6 | 38.6 | 38.6 | 38.4 | 38.8 | 217.6 |
| | BBM-JS | 20.3 | 42.5 | 42.6 | 42.6 | 42.5 | 43.0 | 233.4 |
| | BUPD-JS | 5.9 | 13.9 | 13.9 | 13.9 | 13.9 | 14.0 | 75.5 |
| 7 | $\pi_i=$ | 0.40 | 0.40 | 0.40 | 0.40 | 0.40 | 0.40 | |
| | BBM-NB | 2.0 | 2.0 | 2.0 | 2.0 | 2.0 | 2.0 | 12.0 |
| | MEM | 44.1 | 44.2 | 43.8 | 44.2 | 43.9 | 44.1 | 264.4 |
| | BBM-JS | 48.5 | 48.6 | 48.0 | 48.6 | 48.4 | 48.7 | 290.9 |
| | BUPD-JS | 11.9 | 12.0 | 11.7 | 11.9 | 11.8 | 12.0 | 71.3 |
| 8 | $\pi_i=$ | 0.05 | 0.10 | 0.20 | 0.30 | 0.40 | 0.50 | |
| | BBM-NB | 2.0 | 2.0 | 2.0 | 2.0 | 2.0 | 2.0 | 12.0 |
| | MEM | 29.4 | 32.5 | 35.0 | 33.4 | 28.7 | 22.3 | 181.4 |
| | BBM-JS | 26.1 | 31.6 | 36.4 | 35.6 | 30.7 | 23.8 | 184.2 |
| | BUPD-JS | 10.4 | 14.1 | 17.1 | 16.6 | 13.6 | 9.7 | 81.5 |

**S4. Posterior mean of parameter $s$ for BUPD-JSH in the simulation study**

Table S2 shows the posterior mean of $s$ for BUPD-JSH in scenarios 1–8 in the simulation study of Section 3. BUPD-JSH exhibited a lower posterior mean of $s$ (i.e., widely varying value for each $w_{ij}$) in scenarios with heterogeneous response rates (e.g., scenarios 3–5 and 8) and a higher posterior mean of $s$ (i.e., similar value for each $w_{ij}$) in scenarios with homogeneous response rates (e.g., scenarios 1 and 7). This fluctuation in the posterior mean of $s$ for each scenario improved the accuracy of estimation of $\boldsymbol{w}$ compared with that of BUPD-JS, which is based on Equation (9) (see Figure S1 in Section S3).

**Table S2.** Posterior means of $s$ for BUPD-JSH in scenarios 1–8.

| Scenario | 1 | 2 | 3 | 4 | 5 | 6 | 7 | 8 |
|---|---|---|---|---|---|---|---|---|
| Mean | 12.6 | 7.5 | 4.2 | 3.4 | 3.9 | 6.0 | 9.1 | 3.2 |

## S5. Posterior means of parameters $w_{ij}$ and $M$ for the three proposed methods in the simulation study

Figure S1 shows the posterior mean of $w_{ij}$ for each combination of the two cancer types in BUPD-D, BUPD-JS, and BUPD-JSH in scenarios 1–8 in the simulation study of Section 3. The three proposed methods had almost the same values of $w_{ij} = 1/I(I-1) \approx 0.033$ in scenarios with homogeneity of response rates (e.g., scenarios 1 and 7). In scenarios with heterogeneity of response rates (e.g., scenarios 2–6), each value of $w_{ij}$ under true response rates of $\pi_i = \pi_j$ for BUPD-D, BUPD-JS, and BUPD-JSH was higher than each value of $w_{ij}$ under true response rates of $\pi_i \neq \pi_j$ for these three methods. In scenarios 2 and 6, the values of $w_{ij}$ under $\pi_i = \pi_j$ in BUPD-JS and BUPD-JSH were higher than those in BUPD-D, and the values of $w_{ij}$ under $\pi_i \neq \pi_j$ in BUPD-JS and BUPD-JSH were lower than those in BUPD-D. However, each value of $w_{ij}$ in BUPD-JS was the same as the corresponding value in BUPD-JSH. In scenarios 3–5, the values of $w_{ij}$ under $\pi_i = \pi_j$ in BUPD-JSH were the highest, followed by those in BUPD-JS and BUPD-D, and the values of $w_{ij}$ under $\pi_i \neq \pi_j$ in BUPD-JSH were the lowest, followed by those in BUPD-JS and BUPD-D. Therefore, BUPD-JSH demonstrated the most improved estimation accuracy for $w_{ij}$ among the three proposed methods. This improvement of BUPD-JSH over BUPD-JS was achieved by using a hyperprior on $s$ to estimate the response rate, unlike BUPD-JS, which uses a fixed value of $s = 1$ (see Section 2.2).

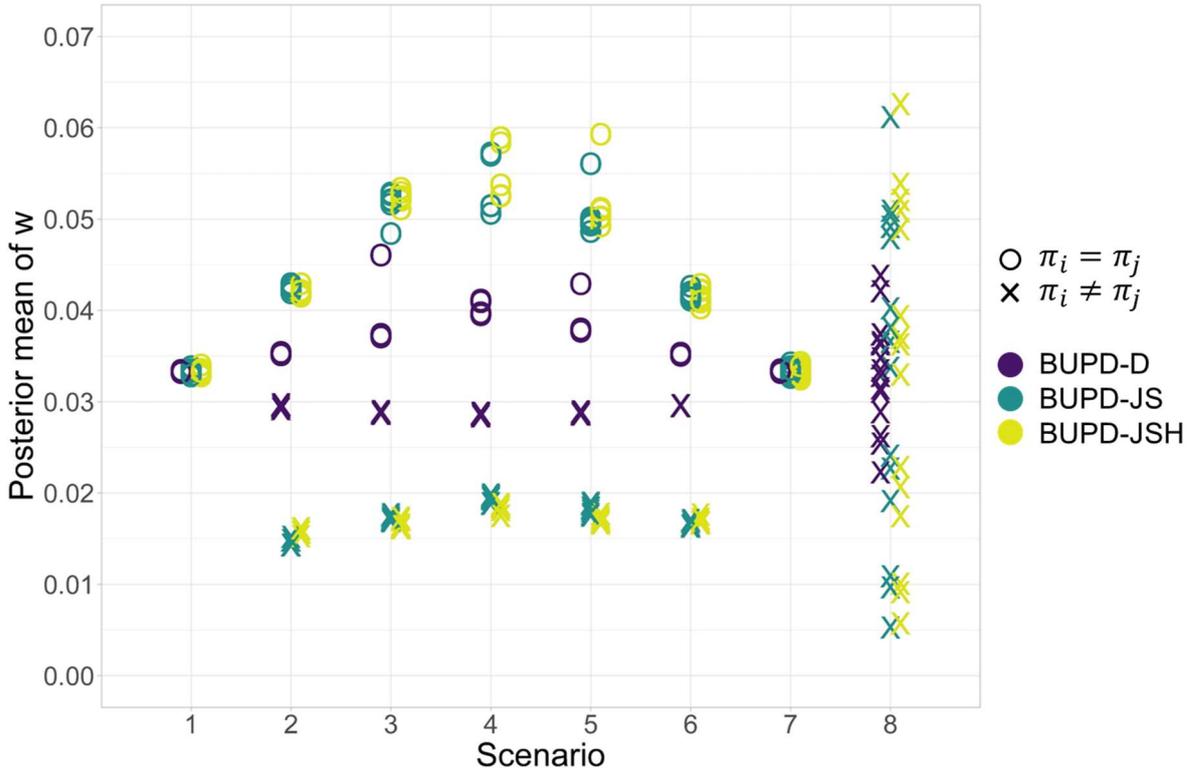

**Figure S2.** Posterior mean of $w_{ij}$ for each combination of two cancer types in BUPD-D, BUPD-JS, and BUPD-JSH in scenarios 1–8. Circular and cross points indicate $w_{ij}$ under true response rates of $\pi_i = \pi_j$ and $\pi_i \neq \pi_j$, respectively.

Table S3 shows the posterior means of $M$ for BUPD-D, BUPD-JS, and BUPD-JSH in scenarios 1–8 in Section 3 of the simulation study. BUPD-D using the hyper-prior on $M$ exhibited a <50 posterior mean of $M$ in scenarios with heterogeneous response rates (e.g., scenarios 2–5 and 8) and >50 posterior mean of $M$ in scenarios with homogeneous response rates (e.g., scenarios 1 and 7). Therefore, BUPD-D reduced the inflation of type 1 error rate by weakening the information borrowing between cancer types, where the true response rates are different. In contrast, the posterior means of $M$ for BUPD-JSH did not vary widely for each scenario and were higher than those for BUPD-D in all eight scenarios. If not considering the heterogeneity in response rates (i.e., $w_{12} = \cdots w_{I(I-1)} = 1/I(I-1)$), reducing the value of $M$ is only one way to weaken the information borrowing between two cancer types with truly different response rates. If $w_{ij}$ varies widely, reducing the value of $M$ is not necessary as it can weakly borrow information between two cancer types with truly different response rates and strongly borrow information between two cancer types with truly equal response rates. Differences in the operating characteristics of $M$ between BUPD-D and BUPD-JSH are closely related to the accuracy of estimation of $w_{ij}$. Compared with BUPD-JS, BUPD-D and BUPD-JSH exhibited lower posterior means of $M$ as UPS-De-JS used a fixed value of $M$. Considering these results for parameters $w_{ij}$ (Figure S1) and $M$ (Table S3), BUPD-JSH reduced the type 1 error rate better than BUPD-D and BUPD-JS in several scenarios (e.g., scenarios 2–6 and 8) by effectively controlling the values of $w_{ij}$ and $M$.

**Table S3.** Posterior means of $M$ for BUPD-D, BUPD-JS, and BUPD-JSH in scenarios 1–8.

| Scenario | 1 | 2 | 3 | 4 | 5 | 6 | 7 | 8 |
|---|---|---|---|---|---|---|---|---|
| BUPD-D | 50.4 | 43.9 | 41.2 | 41.4 | 43.1 | 46.9 | 52.1 | 40.0 |
| BUPD-JS | 72.0 | 72.0 | 72.0 | 72.0 | 72.0 | 72.0 | 72.0 | 72.0 |
| BUPD-JSH | 51.9 | 49.8 | 50.8 | 51.4 | 51.5 | 51.8 | 53.5 | 51.0 |

## S6. Posterior means and 95% equal-tailed credential interval widths of response rates in the simulation study

Figure S3 shows the bias of the posterior mean of the response rates against the true response rates for each cancer type of the eight methods in scenarios 1–8. In all scenarios, BBM-NB and BBM-JS consistently overestimated the response rates because of their use of a $Beta(1,1)$ prior, which pulled the posterior response rates to 50%. BHM, MEM, and BUPD-D exhibited similar biases in most scenarios, although MEM had the largest bias for ineffective cancer types in the scenarios with many effective cancer types (e.g., scenario 6). BUPD-JS and BUPD-JSH demonstrated the smallest biases for ineffective cancer types in all scenarios, followed by EXNEX. For effective cancer types, BBM-JS demonstrated the smallest bias in most scenarios. Although BUPD-JS and BUPD-JSH had larger biases than BBM-NB for effective cancer types in scenarios with few effective cancer types (e.g., scenarios 2–4), they showed smaller biases in the scenarios with many effective cancer types (e.g., scenarios 5 and 6). Across the eight scenarios, BBM-NB, BHM, EXNEX, MEM, BBM-JS, BUPD-D, BUPD-JS, and BUPD-JSH had average biases of 6.1, 4.5, 3.5, 6.0, 8.0, 4.9, 2.4, and 2.6% for ineffective cancer types, and 1.6, –4.0, –2.9, –3.0, –0.4, –3.7, –1.8, and –1.7% for effective cancer types, respectively.

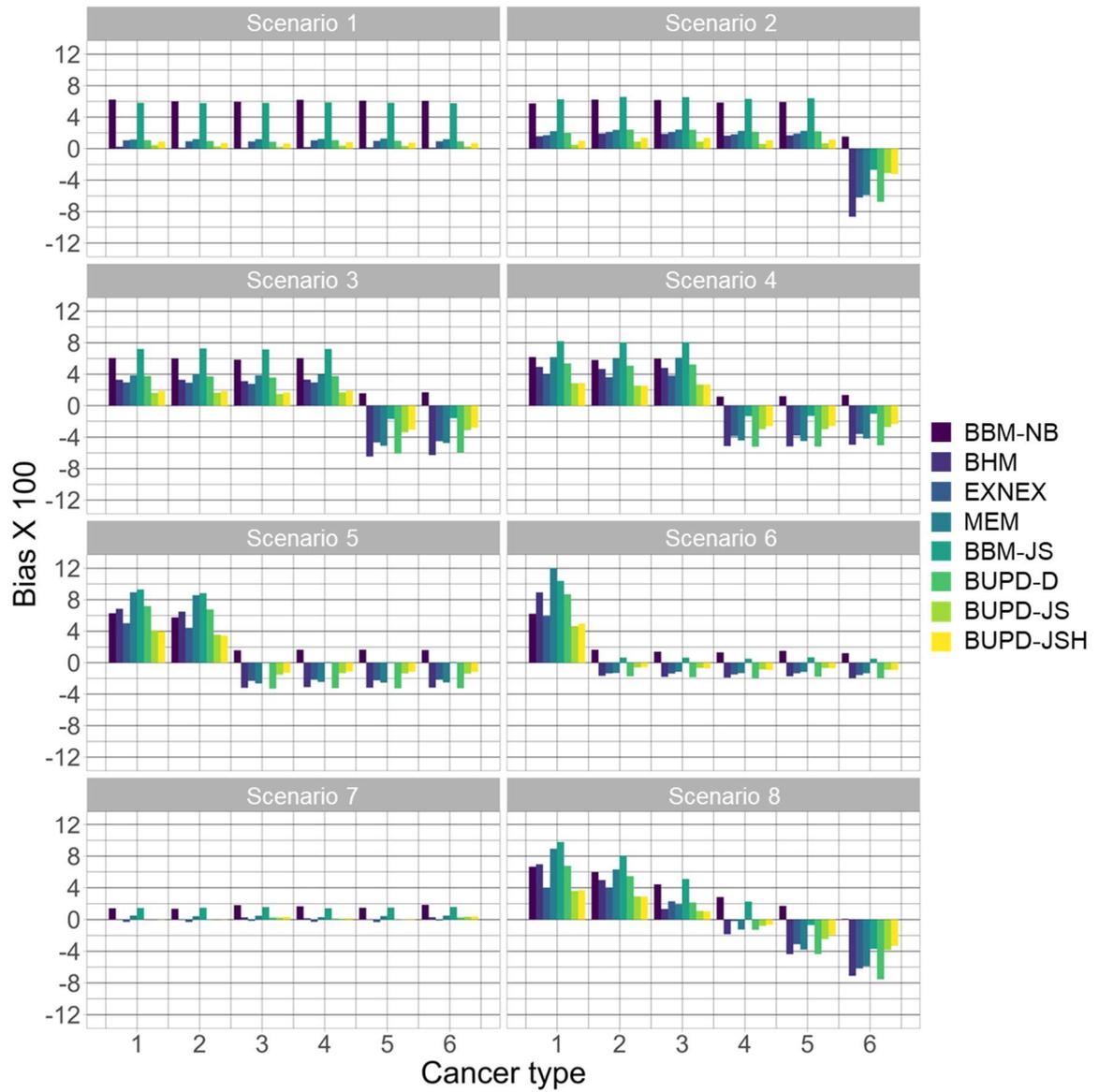

**Figure S3.** Posterior means of the response rates for the eight methods in scenarios 1–8.

Figure S4 shows the 95% equal-tailed credential interval widths (i.e., the difference between the upper and lower limits of the equal-tailed intervals) for the posterior response rates of each cancer type for the eight methods in scenarios 1–8. In all scenarios, BBM-NB exhibited the largest interval width among the eight methods. In contrast, MEM and BBM-JS had the smallest interval widths in all of the scenarios owing to the strong information borrowing between cancer types, as demonstrated in Section S3. BHM and EXNEX had similar interval widths for each scenario. Among the proposed methods, BUPD-JS had the smallest interval width for each cancer type in all scenarios, followed by BUPD-JSH. Across the eight scenarios, BBM-NB, BHM, EXNEX, MEM, BBM-JS, BUPD-D, BUPD-JS, and BUPD-JSH had average 95% equal-tailed credential interval widths of 34.7, 30.0, 30.5, 19.8, 21.5, 30.4, 23.3, and 26.7% for ineffective cancer types, and 47.6, 44.3, 44.0, 29.2, 28.9, 42.5, 36.9, and 41.2% for effective cancer types, respectively.

      In summary, BUPD-JS and BUPD-JSH demonstrated effectiveness in response rate estimation owing to their small competitive average biases and their interval widths. BBM-JS provided a suitable estimation for cancer types with true response rates of approximately 50%. Although MEM showed average interval widths comparable to those of BBM-JS, its larger average biases (i.e., 8.0% for ineffective, and –3.0% for effective cancer types) raise concerns. BBM-NB, BHM, EXNEX, and BUPD-D may be less effective for response rate estimation than BUPD-JS and BUPD-JSH because of their wider average credible intervals, and equal or larger average biases.

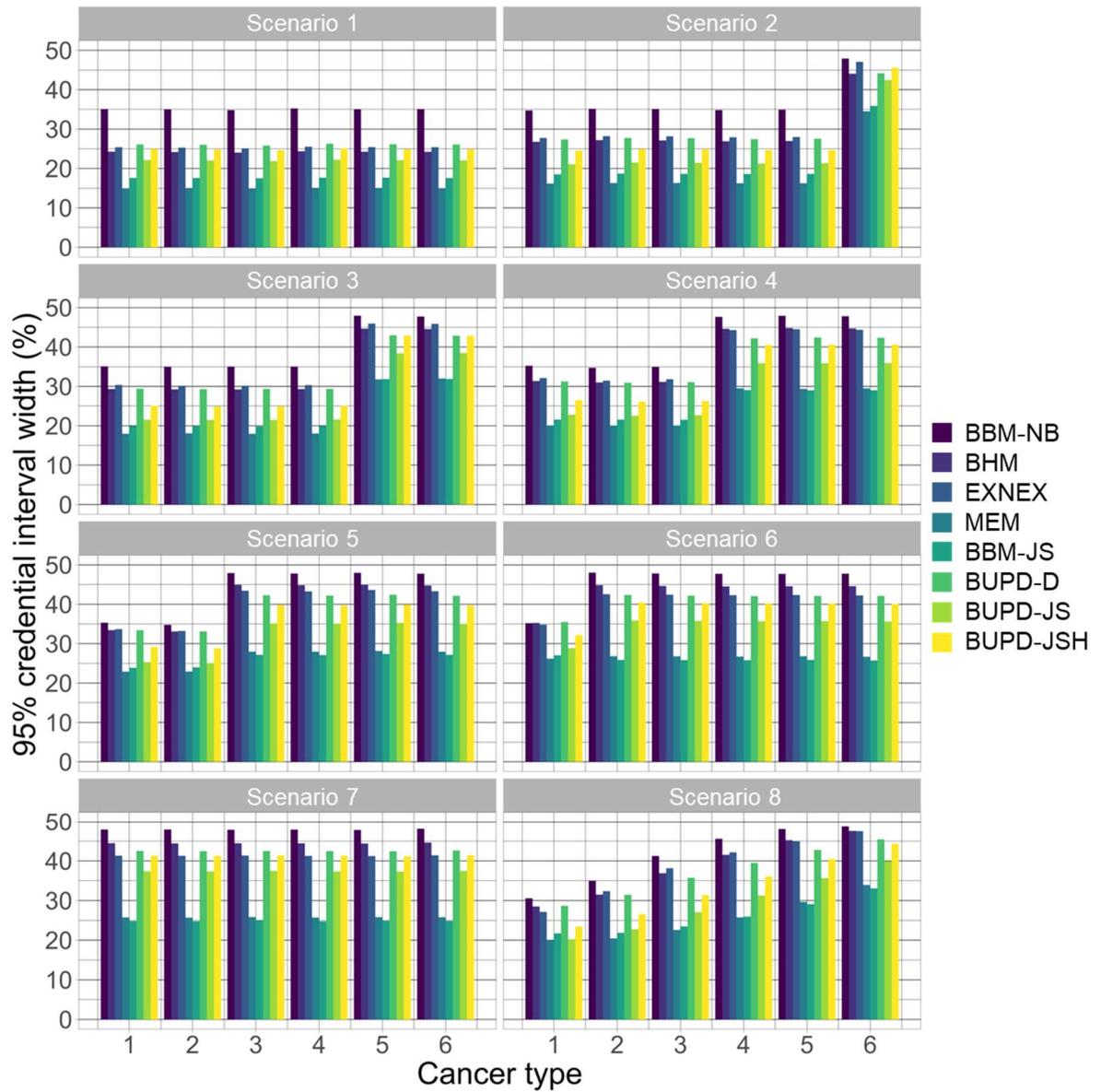

**Figure S4.** 95% equal-tailed credential interval widths of response rates for the eight methods in scenarios 1–8.